\documentclass[12pt]{article}
\usepackage{latexsym}
\usepackage{epsfig,amssymb,euscript}
\usepackage{amsmath}
\usepackage{mathrsfs}
\newsavebox{\ns}
\newsavebox{\dbrane}
\newsavebox{\dbshort}

\def\be{\begin{equation}}
\def\ee{\end{equation}}
\def\bea{\begin{eqnarray}}
\def\eea{\end{eqnarray}}

\def\be{\begin{equation}}
\def\ee{\end{equation}}
\def\ba{\begin{eqnarray}}
\def\ea{\end{eqnarray}}
\newcommand{\vmodk}{V_{5,2}/\Z_k}

\newcommand{\nn}{\nonumber}

\def\Dslash{\,\,{\raise.15ex\hbox{/}\mkern-12mu D}}
\def\Dbarslash{\,\,{\raise.15ex\hbox{/}\mkern-12mu {\bar D}}}
\def\delslash{\,\,{\raise.15ex\hbox{/}\mkern-9mu \partial}}
\def\delbarslash{\,\,{\raise.15ex\hbox{/}\mkern-9mu {\bar\partial}}}
\def\pslash{\,\,{\raise.15ex\hbox{/}\mkern-9mu p}}
\def\calDslash{\,\,{\raise.15ex\hbox{/}\mkern-12mu {\cal D}}}

\newcommand\R{\mathbb{R}}
\newcommand\Z{\mathbb{Z}}

\newcommand\C{\mathbb{C}}

\newcommand\diff{\mathrm{d}}
\newcommand{\de}{\partial}

\newcommand{\dd}{\mathrm{d}}

\newcommand{\ii}{\mathrm{i}}
\newcommand{\vol}{\mathrm{vol}}

\newcommand{\ymodk}{Y_n/\Z_k}
\newcommand{\zflip}{\mathbb{Z}_2^\mathrm{flip}}
\newcommand{\cX}{\mathcal{X}}

\begin{document}
\begin{titlepage}
\begin{center}
\today

\vskip 2 cm 

{\Large \bf AdS$_4/$CFT$_3$ duals from M2-branes at hypersurface\\[2.6mm]
singularities and their deformations}

\vskip 1.5cm
{Dario Martelli$^{1}$ ~~and~~ James Sparks$^{2}$}\\
\vskip 1.2cm

{$^1$\em  Swansea University,\\
Singleton Park, Swansea, SA2 8PP, U.K.}\\

\vskip 0.8cm

{$^2$ \em Mathematical Institute, University of Oxford,\\
24-29 St Giles', Oxford OX1 3LB, U.K.}\\

\end{center}

\vskip 2.5 cm
\begin{abstract}
\noindent
We construct three-dimensional ${\cal N}=2$ Chern-Simons-quiver 
theories which are holographically dual to  the M-theory Freund-Rubin solutions AdS$_4\times V_{5,2}/\Z_k$ 
(with or without torsion $G$-flux), where $V_{5,2}$ is a homogeneous Sasaki-Einstein seven-manifold. 
The global symmetry group of these theories is generically $SU(2)\times U(1)\times U(1)_R$, and they are hence non-toric.
The field theories may be thought of as the $n=2$ member of a family of models, labelled by a positive integer $n$, 
arising on multiple M2-branes at certain hypersurface singularities. 
We describe how these models can be engineered via
generalized Hanany-Witten brane constructions. 
The AdS$_4\times V_{5,2}/\Z_k$ solutions may be deformed to a 
warped geometry $\R^{1,2}\times T^* S^4/\Z_k$, with self-dual $G$-flux through the four-sphere. 
We show  that this solution is dual to
a supersymmetric mass deformation, which precisely modifies the classical moduli space 
of the field theory to the deformed geometry.

\end{abstract}

\end{titlepage}
\pagestyle{plain}
\setcounter{page}{1}
\newcounter{bean}
\baselineskip18pt
\tableofcontents

\section{Introduction}

The work of Bagger and Lambert \cite{BL} (see also \cite{Gustavsson:2007vu}) 
has led to new  insights into the low-energy physics of
M2-branes. In \cite{BL} an explicit three-dimensional ${\cal N}=8$ supersymmetric gauge theory was constructed, a theory which 
 was later shown  to be a Chern-Simons-matter theory \cite{VMS}. Following this work, Aharony, Bergman, Jafferis, and Maldacena
(ABJM) \cite{Aharony:2008ug} have constructed a class of three-dimensional Chern-Simons-quiver theories with 
generically ${\cal N}=6$ supersymmetry (enhanced to ${\cal N}=8$ for Chern-Simons levels $k=1,2$), and argued that
these are holographically dual to the M-theory backgrounds AdS$_4\times S^7/\Z_k$, or their reduction
to Type IIA string theory. This has renewed interest in the AdS$_4$/CFT$_3$ correspondence, opening the way for the construction of many new examples of this duality, in which Chern-Simons theories are believed to play a key role 
\cite{Schwarz:2004yj}. 

An interesting generalization of the ABJM duality is to consider theories with less
 supersymmetry. For example, the case of ${\cal N}=2$ (4 real supercharges) is analogous to
 minimal ${\cal N}=1$ supersymmetry in four dimensions. In the latter case, when the gauge theories are engineered by placing D3-branes at Calabi-Yau singularities the natural candidate holographic duals are given by Type IIB string theory on AdS$_5\times Y^5$, where $Y^5$ is a 
Sasaki-Einstein five-manifold.  It can similarly be argued  
 \cite{Martelli:2008si,Hanany:2008cd,Aganagic:2009zk} that a large class
of Chern-Simons-matter theories should be dual to ${\cal N}=2$ Freund-Rubin vacua
 of M-theory.  This duality, for toric theories, has been studied in 
 many papers -- see, for example, \cite{collective}.

In this paper we will discuss a three-dimensional Chern-Simons-quiver theory that we conjecture to be the holographic dual of M-theory 
on  AdS$_4\times V_{5,2}/\Z_k$, with $N$ units of quantized $G$-flux,
where $V_{5,2}$ (also known as a \emph{Stiefel} manifold) is a homogeneous Sasaki-Einstein 
seven-manifold.  This can be thought of as the near-horizon 
limit of $N$ M2-branes placed at the Calabi-Yau four-fold singularity 
\bea
z_0^2 + z_1^2 + z_2^2 + z_3^2 + z_4^2 \, =\, 0~, \qquad  z_i \in \C~,
\label{thesing}
\eea
which is clearly a generalization of the well-known conifold singulariy in six dimensions. Indeed, Klebanov and Witten 
mentioned this generalization 
in their seminal paper \cite{Klebanov:1998hh}, concluding
with the sentence: ``\emph{We hope it will be possible
to construct a three-dimensional field theory corresponding to M2-branes on (\ref{thesing})}.''  In the present paper we will realize this hope.
We propose\footnote{A different proposal was given in 
\cite{Ceresole:1999zg}. However, this was not based on Chern-Simons theory.}
 that the three-dimensional field theory in question is an ${\cal N}=2$ Chern-Simons-quiver theory with gauge group $U(N)_{k}\times U(N)_{-k}$, generalizing the ABJM model. 
The matter content and superpotential will be presented shortly in section
\ref{sec2}; see Figure \ref{figv} and equation (\ref{superpotential}).

The supergravity solution possesses an $SO(5)\times U(1)_R$ isometry, which reduces to $SU(2)\times U(1)\times U(1)_R$
when we perform a $\Z_k$ quotient analogous to  \cite{Aharony:2008ug} with $k>1$. This is therefore the first example of a 
\emph{non-toric} AdS$_4$/CFT$_3$ duality. In fact there are very few examples of this kind, even in the more developed  
four-dimensional context. The singularity (\ref{thesing}) is the $n=2$ member of a family of $\mathcal{A}_{n-1}$ four-fold singularities, defined by the hypersurface equations 
$X_n = \{ z_0^n + z_1^2 + z_2^2 + z_3^2 + z_4^2 = 0 ~, z_i  \in \C\}$. Thus we are naturally led to 
consider a family of Chern-Simons-quiver theories, labelled by $n$, whose Abelian 
classical moduli spaces are precisely these singularities. Here the $n=1$ model is the ABJM theory of \cite{Aharony:2008ug}. Naively, this suggests that each of these theories will have a large $N$ gravity dual given by AdS$_4\times Y_n$, where $Y_n$ is a Sasaki-Einstein manifold defined by $Y_n = X_n \cap S^9$.
However, the results of \cite{Gauntlett:2006vf} prove that for $n>2$ these Sasaki-Einstein metrics \emph{do not exist}. This means that
the field theories we construct cannot\footnote{We note that it was suggested previously, incorrectly, 
that these singularities lead to AdS$_4$ holographic duals \cite{Gukov:1999ya}.} 
flow to dual conformal fixed points in the IR.
We will review the argument for this in the course of the paper. Nevertheless, we can study these theories in the UV, and in particular we can, and will, discuss their string theory duals in terms of a slight generalization of the Type IIB Hanany-Witten brane configurations \cite{Hanany:1996ie}. This will allow us to derive field theory dualities, in which the ranks 
of the gauge groups change, using the Hanany-Witten brane creation effect. 
We emphasize again that the AdS$_4$ Freund-Rubin solutions exist only in the case $n=1$ (the ABJM theory)  and $n=2$. 

One of the motivations for studying these models is that on 
the gravity side there exists a smooth\footnote{The solution is completely smooth 
 only for $k=1$. For $k>1$ there are orbifold singularities.} supersymmetric 
solution which approaches asymptotically the AdS$_4\times V_{5,2}/\Z_k$ background 
\cite{Cvetic:2000db}.  For $k=1$ this solution is a warped product 
$\R^{1,2}\times T^* S^4$, where $T^* S^4$ denotes the cotangent bundle of $S^4$, and there is a 
self-dual $G$-flux through the $S^4$  zero-section.  In fact, the deformed solution corresponds 
to \emph{deforming}  the hypersurface 
singularity by setting the right hand side of equation 
 (\ref{thesing}) to a non-zero value. 
 This is a complex Calabi-Yau deformation, precisely analogous to the familiar 
 deformation of the  conifold in six dimensions. 
Indeed, superficially this solution looks like  the
 M-theory version of the  Type IIB solution of Klebanov-Strassler \cite{Klebanov:2000hb}.
In the IR the two solutions are precisely analogous; however,  in the UV they 
behave rather differently. 
In particular, the M-theory solution here is asymptotically AdS$_4\times V_{5,2}/\Z_k$, 
without the logarithmic corrections which are a distinctive feature of the 
solutions of  \cite{Klebanov:2000hb,Klebanov:1999rd,Klebanov:2000nc}. 
The topology of the solution at infinity can support only torsion $G$-flux,
but a careful analysis reveals that in fact in the deformed solution this torsion flux is zero.
Thus we are led to conjecture that the theory in the UV is the superconformal 
Chern-Simons-quiver theory above, with \emph{equal} ranks of the two gauge groups. 
We will argue that this solution corresponds to an RG flow triggered
by adding a supersymmetric mass term to the Lagrangian. This was already observed in 
\cite{Herzog:2000rz},  but we will here describe in more detail the deformation in terms of 
the superconformal Chern-Simons  theory.  In particular, we will see how the deformation 
of the field theory modifies the (classical) vacuum moduli space, precisely reproducing
the deformation of the singularity  (\ref{thesing}). 

The plan of the paper is as follows.  In section \ref{sec2} we introduce the Chern-Simons-quiver
field theories:  we compute their classical vacuum moduli spaces and discuss the relation to parent four-dimensional theories. In section \ref{sec3} we discuss  M-theory and Type IIA duals of these Chern-Simons theories. 
In section \ref{branesec} we construct 
 Hanany-Witten brane configurations in Type IIB string theory, and discuss a brane creation effect in these models. In section \ref{defsol} we describe the deformed supergravity solution. 
In section \ref{defsection} we identify this deformed solution in the UV with a specific supersymmetric mass 
deformation of the field theory.  Section \ref{concsec} briefly concludes. 
We relegate some technical details, as well as a different Type IIA dual, to a number of appendices.


\section{Field theories}
\label{sec2}
We begin by describing a family of $d=3$, $\mathcal{N}=2$ Yang-Mills-Chern-Simons quiver theories. 
The family is labelled by a positive integer $n\in\mathbb{N}$, where the $n=1$ 
theory is that of ABJM \cite{Aharony:2008ug}.

\subsection{A family of $d=3$, $\mathcal{N}=2$ Chern-Simons-quiver theories}
\label{csection}
A $d=3$, ${\cal N}=2$ vector multiplet $V$ consists of a gauge field $\mathscr{A}_\mu$,
a scalar field $\sigma$, a two-component Dirac spinor $\chi$, and another scalar field $D$,
all transforming in the adjoint representation of the gauge group.
This is simply the dimensional reduction of the usual $d=4$, ${\cal N}=1$ vector multiplet. 
For the theories of interest, we take the gauge group to be a product $U(N_1)\times U(N_2)$. 
We will therefore have two vector multiplets ${V}_I$, $I=1,2$, with 
corresponding Yang-Mills gauge couplings $g_I$. To the 
usual $\mathcal{N}=2$ Yang-Mills action, we may also add a 
Chern-Simons interaction. This requires specifying the Chern-Simons 
levels $k_I$, $I=1,2$, for the two gauge group factors. These are quantized: for
$U(N_I)$ or $SU(N_I)$ gauge group $k_I \in\Z$ is an integer. 
In this paper we shall only consider the case that 
$k_1=-k_2\equiv k$; for $k_1+k_2\neq 0$ the dual string theory description 
will be in terms of massive Type IIA \cite{alessandro}, which we do not wish to 
consider here. 

\begin{figure}[ht!]
\epsfxsize = 5.5cm
\centerline{\epsfbox{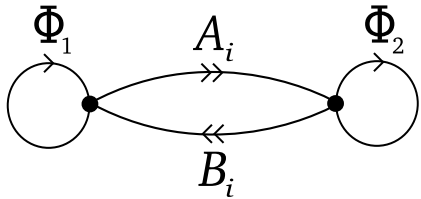}}
\caption{The $\mathcal{A}_1$ quiver.}
\label{figv}
\end{figure}

The matter fields of an $\mathcal{N}=2$ theory are described by chiral multiplets, 
a multiplet consisting of a complex scalar $\phi$, a fermion $\psi$ and
an auxiliary scalar $F$, which may be in an arbitrary representation of the gauge group. 
For the theories of interest, we consider chiral fields
$A_i$, $i=1,2$, transforming 
in the $\mathbf{\bar{N}_1}\otimes \mathbf{N_2}$ representation of $U(N_1)\times U(N_2)$, and bifundamentals 
$B_i$, $i=1,2$, transforming in the conjugate 
$\mathbf{N_1}\otimes \mathbf{\bar{N}_2}$ representation. 
We also introduce chiral fields $\Phi_I$, $I=1,2$, in the adjoint 
representation of $U(N_I)$, respectively.  This gauge and matter content is a quiver gauge theory, 
where the quiver is known as the $\mathcal{A}_1$ quiver. This is shown in 
Figure \ref{figv}.

The total Lagrangian then consists of the four terms (see {\it e.g.} \cite{Gaiotto:2007qi,Martelli:2008si})
\bea
S \, =\, S_{\mathrm{YM}}+ S_{\mathrm{CS}} + S_{\mathrm{matter}} + S_{\mathrm{potential}}~,
\eea
where the bosonic parts of the Chern-Simons and matter Lagrangian are 
\bea
S_{\mathrm{CS}}& = & \sum_{I=1}^2 \frac{k_I}{4\pi}\int  \mathrm{Tr} \,\left( \mathscr{A}_I \wedge \diff \mathscr{A}_I + \frac{2}{3} \mathscr{A}_I\wedge \mathscr{A}_I\wedge \mathscr{A}_I  +
2D_I\sigma_I \right)~,
\label{CSaction}\\
S_{\mathrm{matter}} &=& \sum_a \int \diff^3 x  \mathscr{D}_\mu \bar\phi_a \mathscr{D}^\mu \phi_a - \bar\phi_a\sigma^2 \phi_a + \bar\phi_a D \phi_a \label{mataction}~,
\eea
respectively, where $\phi_a=(A_i,B_i,\Phi_I)$. In (\ref{mataction}), the $\sigma$ and $D$ fields 
act in the appropriate representation on the $\phi_a$ -- see \cite{Gaiotto:2007qi,Martelli:2008si}. The Yang-Mills terms will, at low energies, be irrelevant. 
Finally, the F-term potential is 
\bea
S_{\mathrm{potential}} &=& - \sum_a \int \diff^3 x  \left|\frac{\partial W}{\partial \phi_a}\right|^2 ~,
\label{potaction}
\eea
and we take the following superpotential:
\bea\label{superpotential}
W=\mathrm{Tr}\left[s\left((-1)^n\Phi_1^{n+1}+ \Phi_2^{n+1}\right)+\Phi_2(A_1B_1+A_2B_2)+\Phi_1(B_1A_1+B_2A_2)\right]~.
\eea
Here $n\in\mathbb{N}$ is a positive integer, and $s$ is a complex coupling constant. The superpotential is manifestly invariant under 
an $SU(2)_r$ flavour\footnote{The reason for the subscript $r$ will 
become apparent later. It is \emph{not} to be confused with 
an R-symmetry.} symmetry under which the adjoints $\Phi_I$ 
are singlets and both pairs of  bifundamentals $A_i, B_i$ transform as doublets. 
There is also a $ \zflip $
 symmetry which 
exchanges $\Phi_1\leftrightarrow \Phi_2$, $A_i\leftrightarrow B_i$, 
$s\leftrightarrow (-1)^n s$. 

The case $n=1$ is special, since then the first 
two terms in (\ref{superpotential}) give a mass to the adjoint fields $\Phi_1$, $\Phi_2$. 
At low energy, we may therefore integrate out these fields.
On setting  $s=k/8\pi$, one recovers the ABJM theory with quartic superpotential 
\cite{Aharony:2008ug}
\bea
W_{\mathrm{ABJM}} = \frac{4\pi}{k}(A_1B_2A_2B_1 - A_1B_1A_2B_2)~.
\eea
This theory is in fact superconformal with enhanced manifest ${\cal N}=6$ supersymmetry.
We shall discuss the IR properties of the $n>1$ theories after first discussing their vacuum moduli spaces.

\subsection{Vacuum moduli spaces}
\label{vmsection}

We denote the ranks by $N_1=N+l$, $N_2=N$, and consider the vacuum moduli space of the theory 
$U(N+l)_k\times U(N)_{-k}$. In general there are six F-term equations derived from imposing vanishing of (\ref{potaction}), which is
$\diff W=0$:
\bea\label{Fterms}
B_i\Phi_2 + \Phi_1 B_i&=&0~,\nonumber\\
\Phi_2 A_i + A_i\Phi_1 &=&0~,\nonumber\\
s(n+1)\Phi_2^n + (A_1B_1+A_2B_2)&=&0~,\nonumber\\
s (-1)^n(n+1)\Phi_1^n + (B_1A_1+B_2A_2)&=&0~.
\eea
One must also impose the three-dimensional analogue of the D-term equations \cite{Martelli:2008si}, 
and divide by the gauge symmetry.

It is easier to understand this moduli space in stages, starting with 
the Abelian theory with $k=1$. 
In the $U(1)\times U(1)$ gauge theory, as usual in quiver theories the 
diagonal $U(1)$ decouples (no matter field is charged under it). 
Precisely as in the ABJM theory at Chern-Simons level $k=1$, 
the anti-diagonal $U(1)$, which we denote $U(1)_b$, may be gauged away because of the Chern-Simons interaction. 
Thus the vacuum moduli space, in the Abelian case with $k=1$, is described purely by the set of F-terms 
(\ref{Fterms}). The first four equations are reducible: either $\Phi_1=-\Phi_2$, or else $A_i=B_j=0$ for all $i,j$. 
In the latter case the last two equations imply $\Phi_1=\Phi_2=0$, so this is not a separate branch. 
Thus $\Phi_1=-\Phi_2$ holds in general, and we obtain the single equation for the moduli space
\bea\label{hypersurface}
s(n+1)\Phi_2^n+ A_1B_1+A_2B_2=0~.
\eea
After the  change of coordinates
$z_1=\frac{1}{2}(A_1+B_1)$, $z_2=\frac{\ii}{2}(A_1-B_1)$, $z_3=\frac{1}{2}(A_2+B_2)$, $z_4=\frac{\ii}{2}(A_2-B_2)$, 
$z_0=(s(n+1))^{\tfrac{1}{n}} \Phi_2$, this becomes simply
\bea
X_n\equiv \left\{z_0^n + \sum_{a=1}^4 z_a^2 = 0\right\}~.
\label{nfourfolds}
\eea
For $n=1$ this is indeed just $\C^4$, as one expects since this is the Abelian ABJM theory with $k=1$, which corresponds to
the theory on an M2-brane in flat spacetime. For $n>1$, (\ref{nfourfolds}) instead describes 
an isolated four-fold hypersurface singularity, where the isolated singularity 
is at the origin $\{z_0=z_1=\cdots=z_4=0\}$. This is Calabi-Yau, in the sense 
that away from the singular point there is a global nowhere-zero holomorphic 
$(4,0)$-form. We denote the four-fold singularity by $X$, or $X_n$ when we 
wish to emphasize the $n$-dependence. In particular,  $X_1\cong \C^4$. 
We shall study these varieties in more detail later.

The effect of changing the Chern-Simons levels to $(k,-k)$ 
leads to a discrete quotient of the above vacuum moduli space by 
$\Z_k\subset U(1)_b$  \cite{Aharony:2008ug,Imamura:2008nn,Martelli:2008si}. 
Here by definition the charges of $(A_1,A_2,B_1,B_2)$ 
under $U(1)_b$ are $(1,1,-1,-1)$, while the adjoints are uncharged. 
Thus for general $k$ the Abelian vacuum moduli space is 
$X_n/\Z_k$, where $\Z_k$ acts freely away from the isolated singular 
point. Thus $X_n/\Z_k$ is also an isolated four-fold singularity.

Having understood the moduli space for the $U(1)_k\times U(1)_{-k}$ theory, 
we may now turn to the general non-Abelian $U(N+l)_k\times U(N)_{-k}$ theory. 
The discussion here is similar to that for the ABJM theory in \cite{Aharony:2008ug, Aharony:2008gk}. 
In vacuum, $\Phi_1$, $\sigma_1$ are $(N+l)\times (N+l)$ matrices (with $\sigma_I$ Hermitian), $\Phi_2$, $\sigma_2$ are $N\times N$ matrices, while the $A_i$ and $B_i$ are $N\times (N+l)$ and $(N+l)\times N$ matrices, respectively. 
Note that using the gauge symmetry one may always diagonalize the $\sigma_I$.  The latter are fixed 
by the chiral field VEVs via three-dimensional analogues of the four-dimensional D-term equations 
\cite{Martelli:2008si}, with the $\sigma_I$ playing the role of moment map levels.
If we take \emph{all} 
matrices to be diagonal in the obvious $N\times N$ sub-blocks, so that the chiral fields take the form
\bea
\phi_a^{AB} = \delta^{AB} \phi_a^A~,\qquad A,B=1,\ldots, N~,
\eea
with all other entries zero, then it is simple to see that the scalar potential is zero 
provided the $\phi_a^A$, $A=1,\ldots,N$, satisfy the Abelian equations (the F-terms $\Phi_1^A=-\Phi_2^A$, (\ref{hypersurface}), and 
the D-term equations involving the $\sigma_I^A$). 
It is also straightforward to see from the D-term potential that for \emph{generic} 
$\sigma_I$ (meaning pairwise non-equal eigenvalues), all off-diagonal 
fluctuations about any vacuum in this space of vacua are massive, with the exception 
of fluctuations of $\Phi_1$ in the $l\times l$ sub-block. The diagonal 
ansatz for the fields breaks the gauge symmetry to $U(1)^N\times U(l)\times U(1)^N\times S_N$, 
{\it i.e.} we obtain precisely $N$ copies of the Abelian $N=1$ theory, where the permutation group
$S_N$ permutes the diagonal elements (it is the Weyl group of the diagonal $U(N)$). We also obtain 
a $U(l)_k$ Chern-Simons theory, as in \cite{Aharony:2008gk}, but for general $n$ we also obtain 
a superpotential term $\Psi^{n+1}$, where $\Psi$ is an adjoint under $U(l)$ coming from the 
$l\times l$ sub-block of $\Phi_1$. Classically this has a trivial moduli space, since 
the F-term gives $\Psi=0$. Thus classically we obtain the symmetric product of 
$N$ copies of the Abelian vacuum moduli space, {\it i.e.} $\mathrm{Sym}^N (X_n/\Z_k)$. 

However, as for the ABJM theory, in the quantum theory this moduli space can be lifted. 
In particular, the $U(l)_k$ Chern-Simons theory with an adjoint superpotential 
$\Psi^{n+1}$ has been studied in the literature before -- for a recent 
account, together with a D-brane engineering of this theory, see for example \cite{Armoni:2009vv} and \cite{Niarchos:2008jb}. 
As reviewed in the latter reference, around equation (2.4), the above Chern-Simons theory 
has \emph{no supersymmetric vacuum} unless $0\leq l\leq nk$. This suggests that the above
classical space of vacua is lifted unless this condition on $l$ is obeyed. As we shall see 
later in the paper, this condition is also realized non-trivially in the M-theory dual, and leads 
to a 1-1 matching between the field theories $U(N+l)_k\times U(N)_{-k}$, with $0\leq l <nk$, and the M-theory backgrounds we shall describe 
in section \ref{sec3} (the theories with $l=0$ and $l=nk$ will turn out to be dual to each other under 
a Seiberg-like duality that we derive using the Type IIB brane dual in section \ref{branesec}).

\subsection{IR fixed points}

As mentioned already, for $n=1$ the fields $\Phi_1$, $\Phi_2$ are massive and on integrating these out we recover at low energies 
the ABJM theory. This has $\mathcal{N}=6$ superconformal invariance for general $k\in\Z$. 
For $n>1$ the IR dynamics is rather different. Anticipating much of the discussion that will follow later in section \ref{sec3}, 
we may use the AdS/CFT correspondence to conjecture that the theory with $n=2$ and equal ranks $N_1=N_2=N$ flows to a strongly coupled ${\cal N}=2$ superconformal  fixed point in the IR. The reason for this is that in this case there exists 
a candidate gravity dual: an AdS$_4\times Y_2/\Z_k$ Freund-Rubin solution of eleven-dimensional supergravity, 
where $Y_2$ is a Sasaki-Einstein seven-manifold. More precisely, the four-fold hypersurface singularity 
$X_2$ admits a \emph{conical} Calabi-Yau (Ricci-flat K\"ahler) metric, where the base of the cone 
is described by a homogeneous Sasaki-Einstein metric on $Y_2$ -- we shall discuss this in detail in section \ref{sec3}. 
Notice that, since $W$ has R-charge/scaling dimension precisely 2, all of the fields $\phi_a=(A_i,B_i,\Phi_I)$  
must have R-charge/scaling dimension 2/3 at this fixed point, showing that it is strongly coupled.  
As we shall also see in section \ref{sec3}, more precisely we conjecture this fixed point with equal ranks $N$ to be dual 
to the Freund-Rubin Sasaki-Einstein background with zero internal $G$-flux: as for the ABJM theory \cite{Aharony:2008gk}, more generally it is possible 
to turn on $l$ units of discrete torsion $G$-flux, where in the gravity solution $l$ is an integer mod $nk$,  which is dual to changing the ranks 
to $U(N+l)_k\times U(N)_{-k}$, as discussed at the end of the previous subsection.

On the other hand, it was shown in \cite{Gauntlett:2006vf} that for $n>2$ the natural candidate 
Sasaki-Einstein metrics do not actually exist; that is, the four-fold hypersurface singularities 
$X_n$, for $n>2$, do not have Calabi-Yau cone metrics. This indicates that the corresponding 
field theories cannot flow to conformal fixed points dual to these geometries. Indeed, the 
field theory realization of this was also described in \cite{Gauntlett:2006vf}: if the superpotential 
is (\ref{superpotential}) at the IR fixed point, then the 
gauge invariant chiral primary operators $\mathrm{Tr}\, \Phi_I$ have R-charge/scaling dimension 
$2/(n+1)$; but for $n>2$ this violates the unitarity bound, which requires $\Delta\geq 1/2$, with 
equality only for a free field. It is therefore natural to conjecture that for $n>2$ 
the higher order terms in $\Phi_I$ in (\ref{superpotential}) are irrelevant in the IR, 
and thus $s=0$ at the IR fixed point. If this is the case, then all the theories with $n>2$ flow to the \emph{same} fixed point theory, namely the theory with 
$s=0$.

Consider then setting $s=0$ in $W$ in (\ref{superpotential}). If we also set $k=0$, so that there is no Chern-Simons 
interaction, this is precisely the $\mathcal{A}_1$ quiver gauge theory. 
For equal ranks $N_1=N_2=N$, the latter is well-known to be the low-energy effective theory on
$N$ D2-branes transverse to $\R\times\C\times\C^2/\Z_2$; here $\C^2/\Z_2$, where the generator of $\Z_2$ 
acts via $(z_1,z_2)\mapsto (-z_1,-z_2)$, is precisely the $\mathcal{A}_1$ singularity. 
The latter has an isolated singularity at the origin, where the $N$ D2-branes are placed. 
This may be resolved by blowing up to $\mathcal{O}(-2)\rightarrow\mathbb{CP}^1$ 
(the Eguchi-Hanson manifold). If we wrap $l$ space-filling D4-branes over the 
$\mathbb{CP}^1$ zero-section, the ranks are instead $N_1=N+l$, $N_2=N$. This theory 
has enhanced $\mathcal{N}=4$ supersymmetry. If we now turn on the Chern-Simons coupling 
$k\neq 0$,  the Abelian vacuum moduli space of the resulting theory is easily checked to be 
$\C\times \mathrm{Con}/\Z_k$, where $\mathrm{Con}=\{xy=uv\}\subset \C^4$ denotes the 
conifold three-fold singularity. Since this (non-isolated) four-fold singularity certainly 
admits a Calabi-Yau cone metric, this describes the candidate AdS dual to the IR fixed points of 
the theories with $n>2$. It would be interesting to study this further.

\subsection{Parent $d=4$, $\mathcal{N}=1$ theories and Laufer's resolution}
\label{parent}
As discussed in \cite{Martelli:2008si}, the gauge group, matter content and superpotential 
of a $d=3$, $\mathcal{N}=2$ Chern-Simons matter theory also specify 
a $d=4$, $\mathcal{N}=1$ gauge theory -- one takes the same 
Yang-Mills action, matter kinetic terms and superpotential interaction, 
now defined in $d=4$, and simply discards the Chern-Simons level data 
(since the Chern-Simons interaction doesn't exist in four dimensions). 
This is commonly referred to as the ``parent theory''.
The classical vacuum moduli space of this $d=4$ parent theory is closely 
related to that of the $d=3$ Chern-Simons theory \cite{Martelli:2008si}. 
The string theoretic relation between the two theories was recently elucidated in 
\cite{Aganagic:2009zk}, and we shall make use of this correspondence 
later in the paper. 
The $d=4$ parents of the above theories have been discussed extensively in the 
literature -- in particular, see \cite{Cachazo:2001sg}. We are not interested in the four-dimensional theories 
directly; however, it will be useful to analyse their Abelian vacuum
moduli spaces, and in particular the moduli spaces with a non-zero Fayet-Iliopoulos (FI) parameter 
turned on.

Compared to the $d=3$ Chern-Simons matter theory, the only difference in constructing the Abelian vacuum moduli space of the $d=4$ parent 
is that the $U(1)_b$ gauge symmetry now acts faithfully on the vacuum moduli space. 
The analysis of the F-term equations is identical to that in 
section \ref{vmsection}, and for the Abelian theory with equal ranks $N_1=N_2=1$ we obtain the hypersurface equation (\ref{hypersurface}). 
However, we must also impose the D-term
\bea\label{Dterm}
|A_1|^2 + |A_2|^2 - |B_1|^2 - |B_2|^2 = \zeta~,
\eea
and divide by $U(1)_b$. Here we have introduced
an FI  parameter $\zeta\in\R$ for $U(1)_b$. 

Let us first set $\zeta=0$. 
In this case, the combination of the D-term (\ref{Dterm}) 
and identifying by $U(1)_b$ may be realized holomorphically 
by taking the holomorphic 
quotient by the complexification $\C^*_b$. The charges 
of $(A_1,A_2,B_1,B_2)$ are $(1,1,-1,-1)$, and thus 
the invariant functions on the quotient are 
spanned by $x=A_2B_2$, $y=A_1B_1$, $u=A_1B_2$, 
$v=A_2B_1$. These satisfy the single relation
\bea\label{conifold}
xy=uv~,
\eea
which is the conifold singularity. We must also impose the 
F-term (\ref{hypersurface}), which setting $z_0=(s(n+1))^{\frac{1}{n}} \Phi_2$, as before, reads
\bea\label{hypersurfaceagain}
x + y + z_0^n = 0~.
\eea
Combining (\ref{hypersurfaceagain}) with (\ref{conifold}), and 
again changing variables $u=A_1B_2=\ii w_2 - w_3$, 
$v=A_2B_1 = \ii w_2 + w_3$, $y=A_1B_1=\ii w_1 - w_0^n$, 
$z_0=[s(n+1)]^{1/n} \Phi_2 = 2^{1/n} w_0$ gives the 
three-fold singularity
\bea\label{threefold}
W^0_n\equiv \left\{w_0^{2n} + w_1^2+w_2^2+w_3^2=0\right\}~.
\eea
This is an isolated three-fold singularity, and is again Calabi-Yau in the sense 
that there is a holomorphic volume form on the complement 
of the singular point $\{w_0=w_1=w_2=w_3=0\}$.

Taking the parameter $\zeta\neq 0$ in (\ref{Dterm}), one obtains a 
``small'' resolution  of the singularity $W^0_n$. It is small in the sense 
that the singular point is replaced by a one-dimensional (rather than two-dimensional) complex submanifold --
specifically, a $\mathbb{CP}^1$. 
More precisely, for $\zeta>0$ we obtain a resolution $W^{\zeta}_n\cong W_n^+$, where ``$\cong$'' means biholomorphic, while for 
$\zeta<0$ we obtain a resolution $W^{\zeta}_n\cong W_n^-$. In both cases the ``exceptional'' 
$\mathbb{CP}^1$ has size $|\zeta|$ in the induced K\"ahler metric. Indeed, any 
K\"ahler metric on $W_n^{\zeta}$ will have a K\"ahler class in $H^2(W^{\zeta}_n,\R)\cong\R$, and 
we regard $\zeta$ as specifying this K\"ahler class. 
Both resolutions are also Calabi-Yau, in the sense that there is a holomorphic volume form, and are thus ``crepant''. 

\subsubsection*{More on $W_n^\zeta$}

The end of this section is more technical, and may be skipped on a first reading.

To see why $W^{\zeta}_n$ takes the form described above, recall that the F-term equation (\ref{hypersurface}) 
describes the moduli space in terms of coordinates $(A_1,A_2,B_1,B_2,\Phi_2)$ on 
$\C^5$. Imposing the D-term (\ref{Dterm}) and dividing by 
$U(1)_b$ then gives $\mathrm{Con}_\zeta\times \C$, 
where the resolved conifold $\mathrm{Con}_\zeta$ is obtained from the quotient 
of the $(A_1,A_2,B_1,B_2)$ coordinates, while the VEV of $\Phi_2$ is a 
coordinate on $\C$. In particular, $\zeta>0$ and $\zeta<0$ are related by the 
conifold flop transition. The exceptional $\mathbb{CP}^1$ in the resolved conifold is at $B_1=B_2=0$ for $\zeta>0$, and 
$A_1=A_2=0$ for $\zeta<0$, respectively. The three-fold $W^{\zeta}_n$ is then
embedded in $\mathrm{Con}_\zeta\times \C$ via (\ref{hypersurface}). 
We may also realize the D-term mod $U(1)_b$ as a $\C^*_b$ quotient. 
Strictly speaking, this is 
a geometric invariant theory quotient, and for $\zeta>0$ we need to remove 
the (unstable) points $\{A_1=A_2=0\}$, while for $\zeta<0$ 
we instead remove $\{B_1=B_2=0\}$. Without loss of generality 
we henceforth take $\zeta>0$ (as $\zeta<0$ is just related by a flop), and thus remove 
$\{A_1=A_2=0\}$ from $\C^4$, spanned by $(A_1,A_2,B_1,B_2)$. Define coordinate patches 
$U_i=\{A_i\neq0\}\subset \C^4$, $i=1,2$. These will cover 
the manifold, as $A_1$ and $A_2$ cannot both be zero. On $U_1$ the invariant functions under 
$\C^*_b$ are spanned by 
$x=A_2B_2$, $y=A_1B_1$, $u=A_1B_2$, $v=A_2B_1$, $\xi = 
A_2/A_1$, while 
on $U_2$ the invariant functions are the same 
$x,y,u,v$, but instead $\mu=A_1/A_2$. We then have the relations
\bea
x=u\xi~, \qquad v = y\xi~, \qquad &\mbox{on}& \quad U_1~,\nonumber\\
u=x\mu~, \qquad y = v\mu~, \qquad & \mbox{on}& \quad U_2~.
\eea
It follows that we may coordinatize $U_1$ by 
$(u,y,\xi)$ and $U_2$ by $(x,v,\mu)$, with transition 
functions $(x,v,\mu)=(u\xi,y\xi,1/\xi)$ on the overlap
 $U_1\cap U_2$. This shows explicitly the 
resolved conifold as $\mathcal{O}(-1)\oplus 
\mathcal{O}(-1)\rightarrow\mathbb{CP}^1$, where 
$\xi$ and $\mu$ are coordinates on the two patches 
of the Riemann sphere $\mathbb{CP}^1$,  with $\mu=1/\xi$ on the overlap. 
The poles of the sphere are thus $\mu=0$ and $\xi=0$. 

The three-fold $W_n^+\cong W_{\zeta>0}$ is embedded as a complex 
hypersurface in the
resolved conifold times $\C$. We thus 
introduce patches $H_1$, with coordinates 
$(u,y,\xi,Z_1)$, and $H_2$, with coordinates 
$(x,v,\mu,Z_2)$, where $Z_1=Z_2=\Phi_2$ is the coordinate 
on $\C$. The embedding equation (\ref{hypersurface}) is then simply 
\bea
y=-u\xi-Z_1^n\qquad &\mbox{on}&\quad H_1\nonumber~,\\
x=-v\mu-Z_2^n\qquad &\mbox{on}&\quad H_2~.
\eea
We may thus eliminate $x$ and $y$ and coordinatize $H_1$ by 
$(u,\xi,Z_1)$ and $H_2$ by $(v,\mu,Z_2)$, with transition functions
$(v,\mu,Z_2)=(-\xi Z_1^n - \xi^2 u,1/\xi,Z_1)$ on the overlap
$H_1\cap H_2$. This is precisely the description 
of the small crepant resolution $W_n^+$ of 
$W^0_n$ given by Laufer \cite{Laufer}. One sees explicitly 
the exceptional $\mathbb{CP}^1$ with 
coordinates $\xi,\mu$, and $\mu=1/\xi$ on the overlap.
One also sees that for $n=1$ the normal bundle of 
$\mathbb{CP}^1$ inside $W_n^+$ is $\mathcal{O}(-1)\oplus \mathcal{O}(-1)\rightarrow 
\mathbb{CP}^1$, while for all $n\geq 2$ the normal bundle is instead 
$\mathcal{O}(0)\oplus \mathcal{O}(-2)\rightarrow\mathbb{CP}^1$.


\section{M-theory and Type IIA duals}
\label{sec3}

In this section we discuss M-theory and Type IIA duals to the Chern-Simons-quiver theories of section \ref{csection}. 
We have already shown that the vacuum moduli space of the $U(N+l)_k\times U(N)_{-k}$ theory is $\mathrm{Sym}^N X_n/\Z_k$, 
and this suggests a dual M-theory interpretation in terms of $N$ M2-branes probing the four-fold singularity 
$X_n/\Z_k$. As in \cite{Aharony:2008gk}, we show that the integer $l$, which is constrained to lie in the interval 
$0\leq l \leq nk$ in the field theory, may be identified with turning on $l$ units of torsion $G$-flux 
in the M-theory background. On the gravity side, $l$ is defined only modulo $nk$ -- we will have to wait until 
section \ref{branesec} to see why the $l=0$ field theory is dual to the $l=nk$ theory. 

As already mentioned, only for $n=1$, $n=2$ do the four-fold singularities $X_n$ have Ricci-flat K\"ahler \emph{cone} metrics, 
implying that only in this case do the conformal fixed points of the Chern-Simons-quiver theories have AdS duals of this type; we conjectured that 
for all $n>2$ the theories flow to the \emph{same} fixed point theory in the IR, and that this has a different AdS dual 
description where the Sasaki-Einstein seven-space is the singular link of $\C\times\mathrm{Con}/\Z_k$. 
Although we are interested primarily in the case $n=2$, we retain $n$ throughout this section and study M-theory on 
AdS$_4\times Y_n/\Z_k$, where $Y_n$ is the link of the singularity $X_n$. 
We stress again, however, that the AdS$_4$ solutions of this type exist \emph{only} for $n=1$, $n=2$. 

\subsection{M-theory duals}
\label{mdual}

The discussion of section \ref{vmsection} suggests that the Chern-Simons quivers of section \ref{csection}
should have M-theory duals in terms of M2-branes placed at the four-fold singularities $X_n/\Z_k$
(\ref{nfourfolds}). 
Thus it is natural to conjecture that the IR fixed points of the Chern-Simons quivers, for $n=1$, $n=2$,
are SCFTs dual to the gravity backgrounds AdS$_4\times Y_n/\Z_k$, where $Y_n$ is the base of the cone $X_n$, equipped with 
a Sasaki-Einstein metric. The case 
$n=1$ is just the round metric on $Y_1=S^7$, which is the ABJM model. 
The case $n=2$ leads instead to $Y_2=V_{5,2}$, where $V_{5,2}$ has a homogeneous Sasaki-Einstein metric
that we discuss below. 

Consider the complex cone $X_n$ defined in (\ref{nfourfolds}). We may define the compact 
seven-manifold $Y_n$ via
\bea\label{base}
Y_n \equiv X_n \cap S^9~,
\eea
where $S^9=\{\sum_{i=0}^4 |z_i|^2=1\}\subset \C^5$. For $n=1$ this is simply $Y_1=S^7$, so we focus 
on describing $Y_2$. In this case $X_2$ is a complex quadric, and the vector action of $SO(5)$ on the coordinates 
$z_i$ acts transitively on the seven-manifold $Y_2$, and thus $Y_2=V_{5,2}=SO(5)/SO(3)$ is a coset space. 
$X_2$ is  also invariant under the rescaling $z_i\mapsto \lambda z_i$, for $\lambda\in\C^*$, 
and the quotient $B^6\equiv (X_2\setminus\{0\})/\C^*$ is a compact complex manifold of complex dimension three. 
Equivalently, this may be defined as  $B^6=V_{5,2}/U(1)_R$, where $U(1)_R$ acts on the $z_i$ with charge 
$1$, and thus $B^6 \cong \mathrm{Gr}_{5,2}= SO(5)/SO(3)\times SO(2)$ 
is also a coset space. The space  $\mathrm{Gr}_{5,2}$ is  the Grassmanian of two-planes in  $\R^5$. 

There is an explicit  homogeneous Sasaki-Einstein metric 
on $Y_2=V_{5,2}$, so that the quadric singularity $X_2$ 
has a Ricci-flat K\"ahler cone metric. The Reeb $U(1)$ action is 
precisely the action by $U(1)_R\subset\C^*$ above; thus $V_{5,2}$ is 
a regular Sasaki-Einstein manifold and
the quotient $\mathrm{Gr}_{5,2}$ 
is a homogeneous K\"ahler-Einstein manifold. 
The Sasaki-Einstein metric on $V_{5,2}$ may be written 
explicitly in suitable coordinates \cite{Bergman:2001qi}
\bea
\dd s^2(V_{5,2}) \,=\,  \frac{9}{16}\left[\dd\psi + \frac{1}{2} \cos \alpha (\dd\beta - \cos \theta_1 \dd 
\phi_1 - \cos \theta_2 \dd\phi_2)\right]^2 + \dd s^2 (\mathrm{Gr}_{5,2})~,
\label{expvmetric}
\eea
where 
\bea\label{expgmetric}
\dd s^2 (\mathrm{Gr}_{5,2}) &=&  \frac{3}{32} \Big[ 4\dd\alpha^2 + \sin^2 \alpha (\dd \beta - \cos \theta_1 \dd 
\phi_1 - \cos \theta_2 \dd\phi_2)^2 \nonumber\\
&+&  (1+\cos^2 \alpha ) (\dd\theta_1^2 + \sin^2 \theta_1 \dd\phi_1^2 + \dd\theta_2^2 + \sin^2 \theta_2 \dd\phi_2^2 ) 
\nonumber \\
&+& 2 \sin^2\alpha \cos \beta \sin\theta_1 \sin\theta_2 \dd \phi_1 d\phi_2 - 2\sin^2 \alpha \cos \beta 
\dd\theta_1 \dd \theta_2 \nonumber\\
&+& 2\sin^2 \alpha \sin\beta (\sin\theta_2 \dd\phi_2 \dd\theta_1 + \sin\theta_1 \dd\phi_1 \dd\theta_2) \Big]
\eea
is the homogeneous K\"ahler-Einstein metric on $B^6=\mathrm{Gr}_{5,2}$. 
The ranges of the coordinates are 
\bea
 0 \leq \theta_i \leq \pi~, \quad 0 \leq \phi_i < 2 \pi~, \quad 0 \leq \psi < 2 \pi~,  \quad 0 \leq \alpha \leq \frac{\pi}{2}~,
\quad 0 \leq \beta < 4\pi~.
\eea
The volume of the Sasaki-Einstein metric on $V_{5,2}$ is  \cite{Bergman:2001qi}
\bea 
\mathrm{vol} (V_{5,2}) \,=\, \frac{27}{128}\pi^4~.
\label{totvol}
\eea
Notice the isometry group of the homogeneous metric on $V_{5,2}$ is $SO(5)\times U(1)_R$, and thus in particular this is a \emph{non toric} manifold. 

Thus for $n=1$, $n=2$ we have 
supersymmetric Freund-Rubin backgrounds of eleven-dimensional supergravity of the type
AdS$_4\times Y_n$, with $Y_1=S^7$ and $Y_2=V_{5,2}$. 
The metric and $G$-field take the 
form\footnote{The Einstein metrics 
on AdS$_4$ and $Y_n$ obey $\mathrm{Ric}_{\mathrm{AdS}_4}  =  -3  g_{\mathrm{AdS}_4}$,  
$\mathrm{Ric}_{Y_n}  = 6  g_{Y_n}$, respectively.} 
\bea
\diff s^2 & = & R^2 \left(\frac{1}{4} \diff s^2 (\mathrm{AdS}_4) + \diff s^2(Y_n)\right)~,\nn\\
G & = & \frac{3}{8} R^3 \diff \vol (\mathrm{AdS}_4) ~.
\label{adssolution}
\eea
The  AdS$_4$ radius $R$ is determined by the quantization of the $G$-flux
\bea
N \, = \, \frac{1}{(2\pi l_p)^6} \int_{Y_n} * G~,
\label{gquant}
\eea
where $l_p$ is the eleven-dimensional Planck length, given by
\bea
R^6 \, =\, \frac{(2\pi l_p)^6 N}{6\vol (Y_n)} ~.
\label{bigradius}
\eea
We also note that $\vol(Y_1=S^7)=\pi^4/3$.

Recall that in section \ref{sec2} we introduced an action by the global symmetry group $U(1)_b$. 
Writing the complex cone as $X_n = \left\{z_0^n+ A_1B_1+A_2B_2=0 \right\}$, 
the $U(1)_b$ symmetry acts on $(z_0,A_1,A_2,B_1,B_2)$ with charges $(0,1,1,-1,-1)$. 
This also acts on the base $Y_n$ defined in (\ref{base}), and it is easy 
to see that this is a \emph{free} action, {\it i.e.} there are no fixed 
points on $Y_n$. For both $n=1$, $n=2$, $U(1)_b$ acts isometrically on the 
Sasaki-Einstein metrics. In particular, for $n=2$ this embeds into the isometry group as 
$U(1)_b\cong SO(2)_{\mathrm{diagonal}}\subset SO(4)\subset SO(5)$. 
This is a non-R isometry, and so preserves the Killing spinors on $Y_2=V_{5,2}$. We may thus take a quotient of 
$V_{5,2}$ by $\Z_k\subset U(1)_b$ to obtain a Sasaki-Einstein manifold $\vmodk$ 
with $\pi_1(\vmodk)\cong \Z_k$. Since $SO(4)\cong (SU(2)_l\times SU(2)_r)/\Z_2$, the diagonal $SO(2)$ in $SO(4)$ is $U(1)_b\cong U(1)_l\subset SU(2)_l$. 
Thus the isometry group of the quotient space $\vmodk$ is $SU(2)_r\times U(1)_b \times U(1)_R$. 
This is the manifest global symmetry in the Chern-Simons-quiver theories.

We conjecture that the Chern-Simons-quiver theory $U(N)_k\times U(N)_{-k}$, with matter content 
given by the quiver in  Figure \ref{figv} and superpotential interaction (\ref{superpotential}) with $n=2$,
flows to a conformal fixed point in the IR, and is dual to the above AdS$_4\times Y_2/\Z_k$ M-theory background. 
As evidence for this, we have shown that the moduli space of the field theory agrees with 
the moduli space of $N$ M2-branes probing the cone geometry, and that the isometry group of 
the AdS$_4$ solution precisely matches the global symmetries\footnote{As often happens in AdS$_4/$CFT$_3$, for $k=1$ the isometry group is enhanced.
In particular we have $SO(5)\times U(1)_R$ symmetry, rather than the $SU(2)_r\times U(1)_b\times U(1)_R$ symmetry valid for 
$k>1$. This former symmetry is not manifest in the UV Lagrangian.}  of the field theory. 
Later in sections \ref{multipletsec} and \ref{baryonsec} we shall present a matching 
of various gauge invariant chiral primary operators to supergravity multiplets and 
certain supersymmetric wrapped D-branes, respectively, as further evidence. 
In section \ref{branesec} we will also present a Type IIB brane construction. 

Let us now discuss turning on a torsion $C$-field, corresponding to the addition of fractional branes \cite{Aharony:2008gk}. 
As shown in appendix \ref{apptop}, in general we have $H^4(\ymodk,\Z)\cong\Z_{nk}$, and thus we may turn on
 a torsion\footnote{It is important 
here that the $G$-flux is classified topologically by $H^4(Y,\Z)$, which is true only if the membrane anomaly is zero \cite{membrane}. 
In fact the membrane anomaly always vanishes on any oriented spin seven-manifold.} 
$G$-field, {\it i.e.} a flat, but topologically non-trivial, $G$-flux. Each different choice of such $G$-flux will lead to a physically distinct M-theory background. 
We may equivalently describe this as a (discrete) holonomy for the three-form potential $C$ through the Poincar\'e dual generator $\Sigma^3$ of $H_3(\ymodk,\Z)\cong \Z_{nk}$. Thus
\bea\label{Cperiod}
\frac{1}{(2\pi l_p)^3}\int_{\Sigma^3} C \, =\, \frac{l}{nk}\ \ \mathrm{mod} \ 1~.
\eea 
Since the physical gauge invariant object is a holonomy, the integer $l$ above is only defined modulo $nk$. Equivalently, this 
labels the $G$-flux $[G]=l\in H^4(\ymodk,\Z)\cong \Z_{nk}$.
For each choice of $l$ with $0\leq l < nk$ we therefore have a 1-1 matching of the M-theory backgrounds 
to the field theories with gauge groups $U(N+l)_k\times U(N)_{-k}$. We shall present further 
evidence for matching the $G$-flux to the ranks in this way from the Type IIA dual in section \ref{IIAsec}.

\subsection{Type IIA duals}
\label{redux}

When $k^5\gg  N\gg  k$ the radius of the $U(1)_b$ circle becomes small and a better description is obtained by 
reducing the background along $U(1)_b$ to a Type IIA configuration. 
Since $U(1)_b$ acts freely on $Y_n$, we may define quite generally $M_n=Y_n/U(1)_b$, which is a smooth 
six-manifold. For $n=1$ this gives $M_1=\mathbb{CP}^3$, while for $n>1$ the manifold $M_n$ has the same cohomology groups 
as $\mathbb{CP}^3$, but a cohomology ring that depends on $n$, as shown in appendix \ref{apptop}. 
For $n=2$, $U(1)_b$ is a non-R symmetry, and therefore all supersymmetries are preserved in the quotient 
$V_{5,2}/U(1)_b=M_2$. On the other hand, 
the Type IIA reduction of ${\cal N}=2$ Freund-Rubin backgrounds along the R-symmetry (Reeb vector) direction breaks supersymmetry \cite{Sorokin:1985ap}. 
In particular, we stress that $M_2$ is \emph{different} from the K\"ahler-Einstein six-manifold $\mathrm{Gr}_{5,2}= V_{5,2}/U(1)_R$ 
introduced in section \ref{mdual}. These types of reduction were discussed in \cite{Martelli:2008rt}, and we now recall their essential features. 

To perform the reduction we write the Sasaki-Einstein metric on $Y_n/\Z_k$ as
\bea
\diff s^2 (\ymodk) = \diff s^2(M_n) + \frac{w}{k^2}(\diff \gamma + kP)^2~,
\eea 
where $\gamma $ has $2\pi $ period. We then obtain the following 
Type IIA string-frame metric and fields
\bea
\diff s^2_{\mathrm{st}} \, = \, \sqrt{w}\, \frac{R^3}{k} \left(\frac{1}{4} \diff s^2 (\mathrm{AdS}_4) + \diff s^2(M_n)\right)~,\qquad \quad\\[2mm]
\mathrm{e}^{2\Phi} \, = \,  \frac{R^3}{k^3} w^{3/2}~, \qquad F_4 \, = \,  
\frac{3}{8} R^3 \diff \vol (\mathrm{AdS}_4)~, \qquad F_2 \, = \, k l_s g_s\diff P~,
\eea
where $w$ is a nowhere-zero bounded function on $M_n$ (since $U(1)_b$ acts freely). 
The RR two-form flux has quantized periods, namely 
\bea
\frac{1}{2\pi l_s g_s}\int_{\Sigma^2 } F_2 \, =\, k~.
\label{F2fluxes}
\eea
Here $\Sigma^2\subset M_n$ is the generator\footnote{A detailed discussion of the topology of $M_n$ is contained in appendix \ref{apptop}.} of $H_2(M_n,\Z)\cong \Z$. 
Of course, these supergravity solutions exist only for $n=1$, $n=2$. In the latter case, then more precisely in terms of 
the coordinates in (\ref{expvmetric}), (\ref{expgmetric}) we have that  $\gamma = \phi_2$ and 
\bea
w = \frac{3}{32}\left[1+\tfrac{1}{2}\cos^2\alpha(1+\sin^2\theta_2)\right]~.
\eea

The torsion $C$-field reduces to a flat NS $B_2$-field in Type IIA \cite{Aharony:2008gk} via 
\bea
C = A_3 + B_2\wedge\diff\psi~.
\eea
Here $A_3$ denotes the RR three-form potential, while $\psi$ parametrizes the M-theory circle with
period $2\pi l_sg_s$, where recall that $l_p=l_s g_s^{1/3}$ is the eleven-dimensional Planck length. 
Denoting with $\Omega_2=[\dd P/2\pi]$ the generator of $H^2(M_n,\Z)\cong \Z$, we then have\footnote{The authors of \cite{Aharony:2009fc} argue, for the 
ABJM theory $n=1$, that there is a shift in this $B_2$-field period by $1/2$ (in units of $(2\pi l_s)^2$). Notice that, ordinarily, the $B_2$-field 
period through $\Sigma^2$ would be a modulus, able to take any value in $S^1$ (after taking account of large gauge 
transformations). Since this does not affect 
our discussion, we shall not study this further here.}
\bea
B_2=(2\pi l_s)^2 \frac{l}{kn}\Omega_2~.
\label{thebfield}
\eea
The period of $B_2$ through $\Sigma^2$ is hence
\bea
b \equiv \frac{1}{(2\pi l_s)^2}\int_{\Sigma^2} B_2 = \frac{l}{kn} \ \ \mathrm{mod} \ 1~.
\eea
Again, as for the $C$-field period (\ref{Cperiod}) through $\Sigma^3$, this is only defined modulo 1. 
In Type IIA, this is because large gauge transformations of the $B_2$-field 
change the period $b$ by an integer. 

\subsection{Chiral primaries and their dual supergravity multiplets}
\label{multipletsec}

We now turn to a  discussion of the chiral primary operators of the ${\cal N}=2$ gauge theory with $n=2$, and how they
are realized in the gravity dual. 
In the field theory we can construct chiral primary operators by taking appropriately symmetrized 
gauge-invariant traces of products of fields. 
These operators may be denoted very schematically as
$\mathrm{Tr}\, [\Phi^{n_1}(AB)^{n_2}]$. 
They are invariant under $U(1)_b$, and their dimension at the $n=2$ IR fixed point is $\Delta = 2/3 \cdot  (n_1+2n_2 )$. 
However, because of the presence of monopole operators in three dimensions, 
these do not exhaust the list of all chiral primaries \cite{Aharony:2008ug}. The monopole operator with a single unit of 
magnetic flux in the diagonal $U(1)$ transforms in
 the $(\mathrm{Sym}^k(\mathbf{N}_1), \mathrm{Sym}^k(\bar{\mathbf{N}}_2))$ representation of the gauge group, and  following \cite{Aharony:2008ug} 
we may denote it
as $\mathrm{e}^{\ii \tau}$. Using this we can construct generalized gauge-invariant traces as
\bea
\mathrm{Tr}\, [\Phi^{n_1}(AB)^{n_2} A^{m_1k}B^{m_2k}\mathrm{e}^{\ii (m_1-m_2)\tau}]~,  \quad n_i, m_i \in \mathbb{N}~.
\label{allchirals}
\eea 
It is currently not known how to compute the dimensions of monopole operators in strongly coupled 
${\cal N}=2$ Chern-Simons theories \cite{Benna:2009xd}. However, it is plausible that 
in the present case, as conjectured for the ABJM theory \cite{Aharony:2008ug},  
their scaling dimension is zero. 
Assuming this, the dimensions of the operators (\ref{allchirals}) are then
\bea
\Delta = \frac{2}{3}[n_1+2n_2 + (m_1+m_2)k]~.
\eea

These operators may be matched to a tower of states in the Kaluza-Klein spectrum 
on $V_{5,2}$ derived in \cite{Ceresole:1999zg}.  
Consider first setting $k=1$. The spectrum is arranged into supermultiplets, 
labelled by representations of $Osp(4|2)\times SO(5)\times U(1)_R$. 
When the corresponding dimensions of dual operators
are rational, the multiplets undergo shortening conditions \cite{Ceresole:1984hr}. 
In particular, we see from Table 6 of \cite{Ceresole:1999zg} that a certain vector multiplet 
(``Vector Multiplet II'') becomes
a short \emph{chiral multiplet}, with  components denoted as $(S/\Sigma, \lambda_L, \pi )$. 
These have spins $(0^+,1/2,0^-)$, respectively, and 
dimensions $(\Delta, \Delta+1/2,\Delta+1)$, with 
\bea
\Delta = \frac{2}{3}m~, \qquad m =1,2,\dots~.
\label{kkdimensions}
\eea 
 The lowest component fields then match the  operators  (\ref{allchirals}) with $m=n_1+2n_2+m_1+m_2$.

For $k>1$ only a subsector 
of these states survive the $\Z_k$ 
projection\footnote{The representations that survive the $\Z_k$ projection are 
the singlets in the decomposition of $[m,0]$ 
under $SO(5) \to SU(2)_r \times U(1)_b$.}. This is most easily seen using the equivalence of chiral primary 
harmonics on $V_{5,2}$ to \emph{holomorphic functions} on the Calabi-Yau cone singularity $X_2$ 
\cite{Gauntlett:2006vf}. These can be expanded in monomials of the form $\prod_{i=0}^4 z_i^{s_i}$, for $s_i \in \mathbb{N}$. Using the results of \cite{Gauntlett:2006vf} (see equation (3.22) of this reference) we determine that the R-charges associated to the coordinates\footnote{For general $n$, the would-be 
R-charges are $n/(n+1)$ for the coordinates $z_1,\dots z_4$ and $2/(n+1)$ for 
the coordinate $z_0$. Therefore for $n>3$ the latter violates the unitarity bound $\Delta\geq 1/2$, which geometrically is the Lichnerowicz bound. For $n=3$ it saturates this bound, but one can still argue that the corresponding Sasaki-Einstein metric on $Y_3$ does not exist \cite{Gauntlett:2006vf}.}
 $z_i$ are all equal to $2/3$, which of course agrees with (\ref{kkdimensions}). When $k>1$ it is convenient to change coordinates and write  the singularity as
\bea
z_0^2 + A_1 B_1 + A_2 B_2 \, =\, 0~,
\eea
which diagonalizes the action of $\Z_k \subset U(1)_b$. Recall that 
under $U(1)_b$ these coordinates have charges $(0,1,1,-1,-1)$, respectively.
Thus for $k>1$  a general holomorphic function may be expanded in monomials of the form 
\bea
z_0^{n_1}A^{p_1}B^{p_2}~, \qquad p_1-p_2 = 0~~ \mathrm{mod} ~~k ~, \quad p_i \in \mathbb{N}~.
\eea
These of course match precisley with the operators (\ref{allchirals}), where 
$p_1=n_2+m_1k$, $p_2=n_2+m_2k$.

For later purposes it will be useful to discuss the structure of the chiral multiplets on the gravity side in a little more detail.
The lowest bosonic components $S/\Sigma$ arise from a linear combination of  metric modes and $C$-field modes in AdS$_4$. The top bosonic components $\pi$ come purely from $C$-field modes in the internal directions, namely from certain massive harmonic three-forms on $Y=V_{5,2}$ -- see Table 1 of \cite{Ceresole:1999zg}. 

In the field theory,  a chiral superfield may be written in superspace notation 
as $\Phi = \phi + \theta \psi + \theta^2 F$. 
The component fields have R-charges $(\Delta,\Delta-1,\Delta-2)$ and scaling 
dimensions $(\Delta,\Delta+1/2,\Delta+1)$, respectively. Then  the bosonic 
physical degrees of freedom of a chiral operator of the form
$\mathrm{Tr}\, \Phi^m$ are a scalar $\phi^m$ with dimension $m\Delta$, and a pseudoscalar
$\psi^\alpha \psi_\alpha \phi^{m-2}$ with dimension $m\Delta+1$. In the gravity dual, these 
 are dual to the scalar modes $S/\Sigma$ and the pseudoscalar modes $\pi$, respectively.

\subsection{Baryon-like operators and wrapped branes}
\label{baryonsec}

In this section we briefly discuss
M5-branes wrapped on certain supersymmetric submanifolds in $Y_n/\Z_k$, and their 
Type IIA incarnation 
as D4-branes wrapped on submanifolds in $M_n$. These correspond to certain ``baryonic'' ({\it i.e.} determinant-like) operators
in the field theories. 

A full analysis of the spectrum of baryon-type operators is beyond the scope of this paper. However, we may provide further evidence for the proposed duality by analysing a certain simple set of operators. 
Thus, for the adjoint fields $\Phi_I$ we may consider the
gauge-invariants $\det \Phi_I$, $I=1,2$. Notice that $\Phi_1$ is 
an $(N+l)\times (N+l)$ matrix, while $\Phi_2$ is $N\times N$. We may also define the (in general \emph{non}-gauge-invariant) operators
\bea\label{ABops}
\mathscr{A}_i^{\gamma_1\cdots\gamma_l} &\equiv& \frac{1}{N!} 
\epsilon_{\alpha_1\cdots\alpha_N} A^{\alpha_1}_{i\, \beta_1}\cdots
A^{\alpha_N}_{i\, \beta_N}\epsilon^{\beta_1\cdots\beta_N\gamma_1\cdots\gamma_l}~,\nonumber\\
\mathscr{B}_{i\, \gamma_1\cdots\gamma_l} &\equiv& \frac{1}{N!} 
\epsilon^{\alpha_1\cdots\alpha_N} B_{i\, \alpha_1}^{\beta_1}\cdots
B_{i\, \alpha_N}^{\beta_N}\epsilon_{\beta_1\cdots\beta_N\gamma_1\cdots\gamma_l}~.
\eea
Here $\mathscr{A}_i$ lives in $\Lambda^l \overline{(\mathbf{N+l})}$, 
the $l$th antisymmetric product of the anti-fundamental representation of $U(N+l)$, while $\mathscr{B}_i$ lives in 
$\Lambda^l (\mathbf{N+l})$ \cite{Aharony:2000pp}. These are gauge-invariant only for $l=0$, 
but even in this case one needs to insert an appropriate monopole operator (see \cite{Benna:2009xd,Klebanov:2009sg} 
for a recent discussion of these operators); we will not study this here.  For $l>0$, one can obtain 
gauge-invariant operators by, for example, taking $(N+l)$ copies 
of $\mathscr{A}_i$ and then contracting with $l$ epsilon symbols 
for $U(N+l)$ (with appropriate monopole operators). This situation is 
clearly much more complicated than it is for D3-branes in Type IIB string theory, and deserves further study. However, as for the ABJM 
theory, the operators (\ref{ABops}) can still be matched to  
wrapped branes in the gravity dual, as we shall explain.

In M-theory we may associate these types of operators to M5-branes wrapping supersymmetric submanifolds. More precisely, these are the boundaries of  divisors in the Calabi-Yau cone -- see, {\it e.g.}, the first reference in \cite{collective}. Given the discussion 
of the Abelian moduli space in section \ref{vmsection}, we may 
associate the operators $\det \Phi_I$ with the divisor $\{z_0=0\}$ 
in the Calabi-Yau cone, while $\mathscr{A}_1$ is associated to 
$\{ z_1 =\ii z_2\}$,  $\mathscr{A}_2$ to $\{ z_3 =\ii z_4\}$, 
$\mathscr{B}_1$  to $\{ z_1 =-\ii z_2\}$,  and $\mathscr{B}_2$ to $\{ z_3 =-\ii z_4\}$. This follows by noting that, in the Abelian theory, 
the operators may be regarded as sections of line bundles over 
the Abelian vacuum moduli space; the divisors we have written are then the zeros of these sections. 

Let us consider first the adjoints. Setting $z_0=0$ in $X_n$ gives $\{z_1^2+z_2^2+z_3^2+z_4^2=0\}$, which is a copy of the conifold singularity. 
Thus the boundary $\Sigma_n^{(0)}$ of this divisor is a copy of $T^{1,1}$, for all $n$. Taking the $\Z_k$ quotient, one obtains instead 
$\Sigma_n^{(0)}/\Z_k=T^{1,1}/\Z_k$,
where recall that $\Z_k$ is embedded in the diagonal $SO(2)$ in 
$SO(4)$. For the main case of interest, $n=2$, 
this can be seen explicitly in the polar coordinates of section \ref{mdual}: the five-dimensional 
submanifold $\Sigma_2^{(0)}$ corresponds to setting $\alpha=\beta =0$, and its volume is $\vol (\Sigma_2^{(0)}) =(3\pi)^3/2^5$. 
We may also compute this volume using the 
results of \cite{Bergman:2001qi,Gauntlett:2006vf}. 
This gives the general result
\bea
\vol(\Sigma_n^{(0)})=\frac{(n+1)^3\pi^3}{4n^3}~.
\eea
This is the volume of the submanifold induced by \emph{any} Sasakian metric 
on $Y_n$ with Reeb vector field weights 
$(4/(n+1),2n/(n+1),2n/(n+1),2n/(n+1),2n/(n+1))$. The latter 
are normalized so that the holomorphic $(4,0)$-form on 
the cone has charge $4$. Similarly, one can compute
\bea\label{SEvol}
\vol(Y_n)= \frac{(n+1)^4\pi^4}{48 n^3}~.
\eea
This is then the volume of a Sasaki-Einstein 
metric on $Y_n$ \emph{if it exists}, which is true only for $n=1$, $n=2$. 
 Using the formula for the dimension of the dual operator 
\cite{Gubser:1998fp} 
\bea\label{conformaldim}
\Delta \, =\, \frac{N}{6}\frac{\pi \vol ( \Sigma)}{\vol (Y)}~,
\eea
we obtain in general $\Delta [\det \Phi_I] = 2N/(n+1)$.  
Notice here that, since $\Sigma_n^{(0)}$ is invariant under 
$U(1)_b$, after taking the $\Z_k$ quotient the dependence on $k$ in the numerator and denominator in (\ref{conformaldim}) cancel. 
This result then matches with the conformal dimensions of the adjoints computed 
from the constraint that the superpotential has scaling dimension 2.

However, the above discussion overlooks an important subtlety: we have two 
operators $\det \Phi_1$, $\det \Phi_2$, but only one divisor. 
Moreover, in the case of unequal ranks, $U(N+l)_k\times U(N)_{-k}$, 
one expects $\det \Phi_1$ to have dimension $\Delta\propto N+l$, while $\det \Phi_2$ should have dimension $\Delta\propto N$. 
In the case of D3-branes wrapping supersymmetric three-submanifolds in Sasaki-Einstein five-manifolds, there can also be 
multiple baryonic operators mapping to the same divisor: 
they are distinguished \cite{Franco:2005sm} physically in the gravity dual by having 
different flat worldvolume connections on the wrapped D3-branes. 
Here we have a wrapped M5-brane, and thus one expects the self-dual 
two-form on its worldvolume to play a similar role. 
Notice also that in general in the conformal dimension formula (\ref{conformaldim}) one expects the on-shell M5-brane worldvolume action to appear in the numerator. In general this action depends on both the self-dual 
two-form and the pull-back of the $C$-field, reducing 
simply to the volume of $\Sigma$ when both are zero. 
Of course, $l\neq0$ corresponds in the gravity dual to having 
a non-zero flat $C$-field. Similarly, in the Type IIA dual 
picture that we discuss below these are wrapped D4-branes, whose 
conformal dimensions should be related to the on-shell 
Dirac-Born-Infeld action, including the $B_2$-field (\ref{thebfield}). We shall not 
investigate this further here, but instead leave it for future work. 

The remaining four dibaryon operators in (\ref{ABops}) correspond to the same type of submanifold; hence,  
without loss of generality, we shall study the $\mathscr{A}_1$ operator.
The locus $\{z_1=\ii z_2\}$ in the Calabi-Yau cone $X_n$ cuts out a \emph{singular} subvariety for general $n$: 
clearly, $z_1$ may take any value in $\C$, but the remaining 
defining equation of $X_n$ implies that $z_0^n+z_3^2+z_4^2=0$, which is a copy of 
the $\mathcal{A}_{n-1}$ singularity. Thus the divisor of interest
is $\C\times (\C^2/\Z_n)$, and the intersection with 
$Y_n$ is then a copy of the \emph{singular} space $\Sigma_n^{(1)}=S^5/\Z_n$. On the other hand, the $\Z_k$ quotient acts freely on 
$\Sigma_n^{(1)}$.
The volume may again be computed from the character formula
\cite{Gauntlett:2006vf}, giving
\bea
\vol(\Sigma_n^{(1)})=\frac{(n+1)^3\pi^3}{8n^2}~,
\eea
and hence conformal dimension $\Delta[\mathscr{A}_i]=nN/(n+1)$. 
Again, notice this precisely matches the scaling dimensions 
of the fields $A_i$ obtained by imposing that the superpotential has 
scaling dimension 2. 

It is instructive to also consider the 
reduction to Type IIA. The wrapped M5-branes above then become D4-branes wrapped on four-dimensional 
subspaces $\Sigma_n^{(i)}/U(1)_b$. 
Since the quotient by $U(1)_b$ does not break supersymmetry of the background, we expect that the four-dimensional submanifolds here will also be supersymmetric; however we have not checked the kappa-symmetry of the wrapped D4-branes explicitly. 

The reduction of $\Sigma_n^{(0)}$
is diffeomorphic to $S^2\times S^2$. More interesting 
is the reduction of the (singular) $\Sigma_n^{(1)}$ subspaces, 
corresponding to the dibaryonic operators (\ref{ABops}) 
with $l$ uncontracted indices. The latter dependence on $l$ may be understood 
by analysing a certain tadpole in Type IIA, as for the ABJM 
theory. 
To discuss  the reduction to Type IIA, it is more convenient to 
use the coordinates $A_i$, $B_i$. The divisor corresponding to 
the $\mathscr{A}_{1}$ operator is then simply $\{z_1=\ii z_2\}=\{A_1=0\}$. The group $U(1)_b$ acts with charge $-1$ on the coordinate
$B_1$, and charges $(1,-1)$  on $(A_2,B_2)$.  
The $\mathcal{A}_{n-1}$ singularity in these coordinates is $z_0^n+A_2B_2=0$. 
Denoting by $u_1,u_2$  standard coordinates  on $\C^2$ under which $\Z_n$ acts as 
$(\mathrm{e}^{{2\pi i}/{n}},\mathrm{e}^{{-2\pi i}/{n}})$, then the invariant functions under $\Z_n$ are 
$A_2=u_1^n$, $B_2=u_2^n$ and $z_0=\mathrm{e}^{i\pi/n} u_1u_2$, from which one 
sees explicitly that $A_2B_2=-z_0^n$. 
Thus $U(1)_b$ acts with weights $(1/n,-1/n)$ 
on the coordinates $(u_1,u_2)$. 
This implies that the quotient is topologically
$\Sigma_n^{(1)}/U(1)_b=(S^5/\Z_n)/U(1)_b\cong \mathbb{WCP}^2_{[n,1,1]}$. The latter is the subspace on which the D4-brane 
is wrapped. It has an isolated $\Z_n$ orbifold singularity 
at the image of $A_2=B_2=0$, which lifts to the 
$\mathcal{A}_{n-1}$ singularity. A simple topological
description of $\mathbb{WCP}^2_{[n,1,1]}$ is 
to take $\mathcal{O}(n)\rightarrow\mathbb{CP}^1$, and then 
collapse the boundary, which is $S^3/\Z_n$, to a point. 
The latter is then the isolated singularity. 
Conversely, the image of $B_1=0$ is a smooth two-sphere which lifts to the $S^3/\Z_n$ link of the $\mathcal{A}_{n-1}$ singularity. 
Thus in general the integral of 
 $F_2/(2\pi l_s g_s)$ 
over this $S^2$ in $\mathbb{WCP}^2_{[n,1,1]}$ is equal to 
$nk$. 

Now, from appendix \ref{apptop} we have that $H_4(M_n,\Z)\cong\Z$. Call the generator $\Sigma^4$.
It is also shown in this appendix that 
the integral of the square of $\Omega_2=1\in H^2(M_n,\Z)\cong\Z$ 
over $\Sigma^4$ is equal to $n$. Now, in general also
$[F_2/2\pi l_sg_s]=k\Omega_2$, and since the first Chern class of 
$\mathcal{O}(n)\rightarrow \mathbb{CP}^1$ is $n$, it follows 
that the integral of the pull-back of $\Omega_2\wedge\Omega_2$ over 
$\mathbb{WCP}^2_{[n,1,1]}$ is equal to $n^2/n=n$. 
This implies that the copy of $\mathbb{WCP}^2_{[n,1,1]}$ on which the BPS D4-brane 
is wrapped is a (singular) representative of the four-cycle $\Sigma^4$ in the smooth six-manifold $M_n$. 

Consider now the Wess-Zumino couplings on the D4-brane wrapped on 
$\mathbb{WCP}^2_{[n,1,1]}$. Due to the presence of the $B_2$-field (\ref{thebfield}), we obtain\footnote{This assumes that 
the worldvolume gauge field flux on $\Sigma^4$ is zero. In fact for \emph{odd} $n$, the smooth locus 
of the wrapped submanifold $\Sigma^4=\mathbb{WCP}^2_{[n,1,1]}$ is not spin, and thus one must turn on a 
$1/2$-integral worldvolume gauge field flux to cancel the resulting Freed-Witten anomaly. 
This is related to the $1/2$-integral shift of $B_2$ (in the case $n=1$) in footnote 9, which cancels this. 
In our case of interest, $n=2$, there is no such shift.} the term
\bea\label{tadypole}
\frac{1}{(2\pi)^4 l_s^5}\int_{\R_{\mathrm{time}}} A\cdot  \int_{\Sigma_4} B_2\wedge F_2  
= l \cdot \frac{g_s}{2\pi l_s^2} \int_{\R_{\mathrm{time}}} A~.
\eea
Here we have performed the calculation
\bea
\int_{\Sigma^4} \frac{l}{nk}\Omega_2 \wedge k\Omega_2 = l~.
\eea
The Wess-Zumino coupling thus induces a 
tadpole for the worldvolume gauge field $A$. 
To cancel this tadpole requires that $l$ fundamental strings end on the D4-brane. In the field theory this corresponds to the fact that the dibaryon operators (\ref{ABops}) have precisely $l$ uncontracted indices  \cite{Aharony:2008gk}. 

The alert reader will notice an important subtlety in this argument: 
in the gravity solution $l$ is defined only modulo $nk$, while in the field theory $0\leq l\leq nk$. In particular, when one states that the 
tadpole requires $l$ fundamental strings to end on the D4-brane, this is only true modulo $nk$. Thus, it must be that $nk$ fundamental strings are physically equivalent to none. In fact this is easy to see in the M-theory lift. The strings lift to $nk$ M2-branes ending on the M5-brane. More precisely, the end of the M2-branes wrap
the M-theory circle that is a smooth $S^1$ in $\Sigma_n^{(1)}$, together with the time direction in AdS$_4$. If we remove the singular locus from $\Sigma_n^{(1)}$, which is a copy of $S^1$, 
we obtain a smooth manifold with fundmental group $\Z_{nk}$ -- removing the singular locus is sensible, since the supergravity approximation will break down near to this locus. This result implies that $nk$ M2-branes ending on the M5-brane can ``slip off'', since $nk$ copies of the circle that they wrap are contractible on the M5-brane worldvolume. This matches nicely with the fact that this is equivalent, via (\ref{tadypole}), to a large gauge transformation of the $B_2$-field.

\subsection{Type IIA derivation of the Chern-Simons theories}\label{IIAsec}

There is a different way of thinking about the 
Type IIA backgrounds discussed in section \ref{redux}, which we explain in this section. 
This demonstrates rather directly  the relationship with the ``parent"  four-dimensional field theories, and elucidates the stringy origin of the Chern-Simons-quiver theories. We will also need the present discussion to derive a Type IIB Hanany-Witten-like brane configuration in the next section. 

\begin{figure}[ht!]
\epsfxsize = 15cm
\centerline{\epsfbox{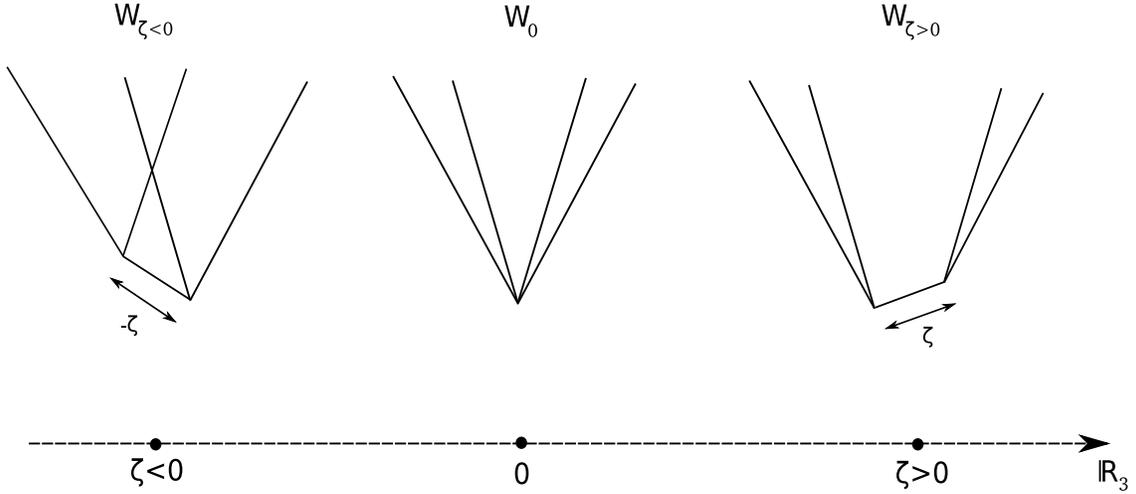}}
\caption{The Type IIA reduction of M-theory on $X/\Z_k$ on $U(1)_b$ is $C(M_n)$. 
This geometry may also be 
viewed as a fibration of $W^{\zeta}_n$ over the $\R_3$ direction, where the size $|\zeta|$ of the 
exceptional $\mathbb{CP}^1$ depends on the position in $\R_3$. In particular, the conical singularity 
of $C(M_n)$ is the conical singularity of $W^0_n$ above the origin in $\R_3$. 
The above schematic picture 
would be precisely the toric diagram in the case $n=1$ (for $n>1$ the geometry is not toric). }
\label{figIIAgeometry}
\end{figure}

We begin by considering the 
geometry $\R^{1,2}\times X_n/\Z_k$ in M-theory, where $X_n$ is the cone singularity (\ref{nfourfolds}), together with 
$N$ spacefilling M2-branes. The $U(1)_b$ circle acts freely away from the cone point, and thus we can 
reduce to a Type IIA geometry $\R^{1,2}\times C(M_n)$, with $k$ units of RR two-form flux through the 
generator of $H_2(M_n,\Z)\cong\Z$. In this picture we have $N$ spacefilling D2-branes.
However, we may  instead take the \emph{K\"ahler quotient} of 
$X_n/\Z_k$ by $U(1)_b$, at level $\zeta\in\R$, to obtain precisely the three-fold $W^{\zeta}_n$ introduced in 
section \ref{parent}. For $\zeta=0$, recall this is the affine three-fold given by (\ref{threefold}), while
for $\zeta\neq0$ one instead obtains Laufer's small resolution of this singularity, which has a blown-up 
$\mathbb{CP}^1$ of size $|\zeta|$. The latter is the Abelian vacuum moduli space of the four-dimensional 
parent theory, as discussed in section \ref{parent}. This picture describes the seven-dimensional space $C(M_n)$ as 
a fibration of $W_n^\zeta$ over the real line $\R$ that parametrizes the moment map level $\zeta$, as 
shown in Figure \ref{figIIAgeometry}.

Indeed, we can instead consider starting with Type IIA on 
$\R^{1,2}\times \R_3\times W^0_n$, where we have labelled $\R=\R_3$ for later convenience, with $N$ 
spacefilling D2-branes. Here $W^0_n$ should of course be equipped with some kind of Calabi-Yau metric, 
although we note that from \cite{Gauntlett:2006vf} it does not admit a \emph{conical} Calabi-Yau metric 
for $n>1$ ($n=1$ is the conifold). We might imagine $W^0_n$ as modelling a local singularity in a 
compact Calabi-Yau manifold, in which case the Calabi-Yau metric here would in any case be incomplete. 
If we now T-dualize along the (compactified) $\R_3$ direction, then we precisely obtain the 
Type IIB string theory set-up yielding the four-dimensional parent theory. 
We may also replace the singular three-fold by its crepant resolution $W^\zeta_n$, thinking of 
$\zeta$ as parametrizing the period of the K\"ahler form through the exceptional $\mathbb{CP}^1$. 
We may then turn on 
$k$ units of RR two-form flux through this $\mathbb{CP}^1$, although in order to 
preserve supersymmetry it is necessary to also fibre 
the size of the $\mathbb{CP}^1$ over the $\R_3$ direction -- this may be seen by appealing to 
the reduction of the M-theory solution above. Thus we identify $\R_3 \cong \{\zeta \in \R \}$.
If $\mu_b$ denotes the moment map for $U(1)_b$, so that 
 $\mu_b:X_n/\Z_k\rightarrow \R_3$, then notice that the inverse 
image of $\zeta\in\R_3$ is $\mu_b^{-1}(\zeta)=W^{\zeta}_n$, so that in particular 
the cone geometry appears at the origin in $\R_3$. 
By construction, the RR two-form flux may then be identified with the first Chern class 
$c_1 \in H^2(W^{\zeta}_n,\Z) $ of the $U(1)_b$ M-theory circle bundle. One can then compute that 
\bea\label{kflux}
\frac{1}{2\pi l_s g_s}\int_{\C P^1} F_2 \, =\, k ~.
\eea 

As explained in \cite{Aganagic:2009zk}, the above picture leads to a physical relation between the 
parent theory and the Chern-Simons theory. If we have $N$ spacefilling D2-branes together with 
$l$ fractional D4-branes wrapping the (collapsed) $\mathbb{CP}^1$ in $W^0_n$, the resulting 
gauge theory is precisely the $\mathcal{A}_1$ quiver theory with superpotential (\ref{superpotential}),  
with gauge group $U(N+l)\times U(N)$ -- this is discussed, for example, in \cite{Cachazo:2001sg}. 
The key result in \cite{Aganagic:2009zk} is that the addition of the $k$ units of RR two-form flux 
through the $\mathbb{CP}^1$ then induces a Chern-Simons interaction with levels $(k,-k)$ for the 
two nodes, respectively, via the Wess-Zumino terms on the fractional branes. This leads to a Type IIA string theory \emph{derivation} of our Chern-Simons-quiver theories, starting with the geometric engineering of the parent theory. Also notice that the $l$ fractional D4-branes, wrapped on the collapsed $\mathbb{CP}^1$, will 
lift to $l$ fractional M5-branes -- since the M5-brane is a magnetic source for the $G$-field, it is thus 
natural to identify the $l$ units of torsion $G$-flux with the $l$ fractional M5-branes. 
Indeed, more precisely, a copy of the exceptional $\mathbb{CP}^1$ at $\zeta>0$ in Figure \ref{figIIAgeometry} is the generator 
of $H_2(M_n,\Z)\cong\Z$, and this lifts to the generator $\Sigma^3$ of $H_3(Y_n/\Z_k,\Z)\cong\Z_{nk}$, as shown in 
appendix \ref{apptop}. Thus $l$ fractional D4-branes wrapped on the $\mathbb{CP}^1$ lift to $l$ 
fractional M5-branes wrapped on $\Sigma^3$. The latter is then Poincar\'e dual to $l$ units of 
torsion $G$-flux. 


\section{Type IIB brane configurations}
\label{branesec}

In this section we derive a 
Hanany-Witten-like brane configuration in Type IIB string theory. This takes the usual form 
of D3-branes (wrapped on a circle) suspended between 5-branes, except that for $n>1$ the 5-branes are embedded non-trivially 
in spacetime; specifically, they are wrapped on holomorphic curves.
This will allow us to understand further aspects of the proposed duality, and also derive a 
field theory duality via a brane creation effect. The reader whose main interest is the deformed $n=2$ supergravity 
solution may wish to skip ahead to section \ref{defsol}.

\subsection{T-duality to Type IIB: $k=0$}

We begin with the Type IIA background of 
 $\R^{1,2}\times \R_3\times W^{\zeta}_n$, with zero RR flux, discussed at the end of the previous section.  
Here we have included a K\"ahler class
$\zeta\in\R$, which is a free parameter, so that for $\zeta\neq0$ $W^{\zeta}_n$ is a smooth non-compact K\"ahler 
manifold.

For $\zeta=0$, we are considering the singular three-fold 
$W^0_n$. We rewrite the defining equation (\ref{threefold}) as
\bea
W^0_n = \{w_0^{2n}+w_1^2-uv=0\}\subset \C^4~,
\eea
where as before $u=\ii w_2- w_3$, $v=\ii w_2+ w_3$. 
We may then consider performing a T-duality along 
$U(1)\equiv U(1)_6$ that acts with charge 
1 on $u$ and charge $-1$ on $v$.  
We may also consider the K\"ahler quotient by $U(1)_6$, 
with moment map $\mu_6=|u|^2-|v|^2$, 
which maps $\mu_6:W^0_n\rightarrow \R\equiv \R_7$, where 
we have introduced the subscript 7 to distinguish this 
copy of $\R$ from $\R_3$ above. It follows that 
$\{\C^2=\langle u, v\rangle\}//U(1)_6\cong\C$, for any value of $\mu_6$, 
and hence similarly $W^0_n//U(1)_6\cong \C^2$. 
Indeed, the defining equation of 
$W^0_n$ is then $w_0^{2n}+w_1^2=w$, 
where $w=uv$ is the coordinate on 
$\C=\C^2/\C_6^*$. We may thus eliminate the 
coordinate $w$ to see that $W^0_n//U(1)_6
\cong \C^2$, spanned by the coordinates $w_0$, $w_1$, for 
any value of the moment map.
It follows 
that $W^0_n/U(1)_6$ is a $\C^2$ fibration over 
$\R_7$, and thus $W^0_n/U(1)_6\cong\R_7\times\C^2\cong \R^5$. 

\begin{figure}[ht!]
\epsfxsize = 11cm
\centerline{\epsfbox{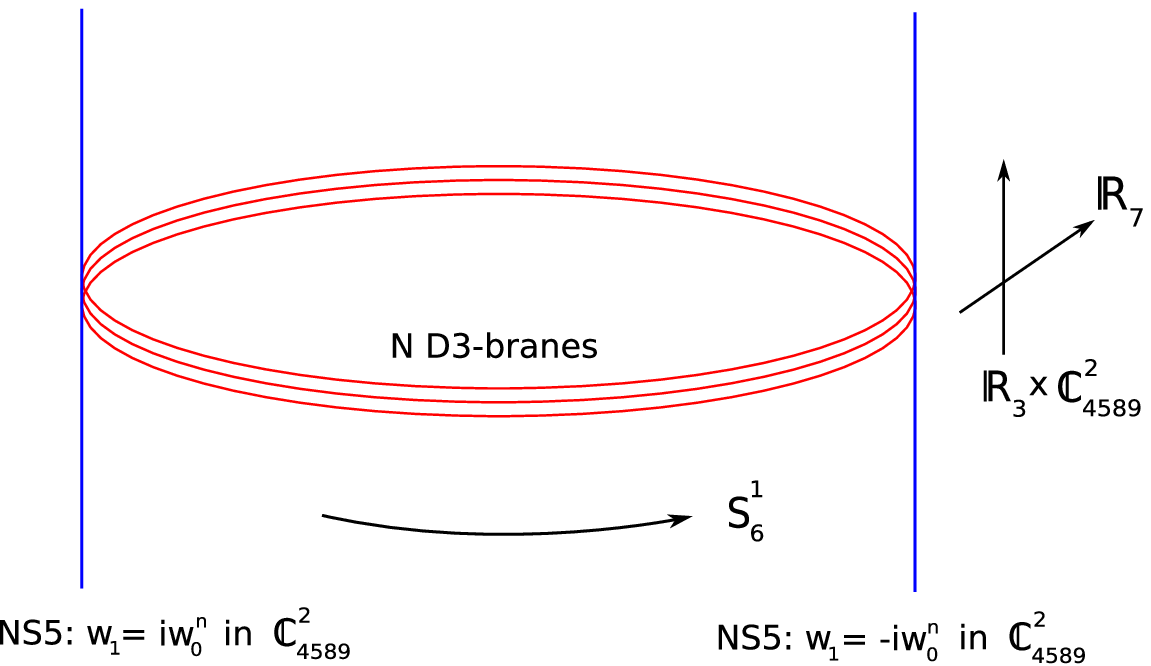}}
\caption{The Type IIB brane dual of the Type IIA background $\R^{1,2}_{012}\times\R_3\times W^0_n$ with 
$N$ spacefilling D2-branes. The Type IIB spacetime is flat: $\R^{1,2}_{012}\times\R_3\times S^1_6\times\R_7\times\C^2_{4589}$. 
There are $N$ D3-branes filling the $\R^{1,2}_{012}$ directions and 
wrapping the $S^1_6$ circle; they are at the origin in $\R_3$, $\R_7$ and $\C^2_{4589}$. 
There are two NS5-branes that are spacefilling in $\R^{1,2}_{012}$ and 
separated by a distance in the $S^1_6$ circle that is given by the 
period of $B_2$ through the collapsed $\mathbb{CP}^1$ in the 
T-dual three-fold geometry $W^0_n$; they both sit at the origin in $\R_7$, fill the $\R_3$ direction, and wrap 
the holomorphic curves $w_1=\pm \ii w_0^n$, respectively, 
in $\C^2_{4589}$ with complex coordinates $w_0,w_1$. These curves intersect 
at the origin $w_0=w_1=0$. $n=1$ is the standard Hanany-Witten brane configuration for 
the conifold singularity, where the NS5-branes are linearly embedded. }
\label{figbranesk0}
\end{figure}

There are, however, fixed points of $U(1)_6$.
If we peform a T-duality along $U(1)_6$, 
the above shows that the T-dual spacetime 
is $\R^{1,2}\times\R_3\times S^1_6\times 
\R_7\times\C^2$, where $S^1_6$ is the $U(1)_6$ 
circle after performing the T-duality. 
However, there are codimension four fixed 
point sets of $U(1)_6$, where the 
action on the normal fibre is the standard Hopf action 
on $\R^4$. These become NS5-branes in the T-dual
Type IIB picture. The fixed locus here is 
$u=v=0$, which is the origin in the moment map 
direction $\R_7$. In the $\C^2$ direction they cut out the locus $w_0^{2n}=-w_1^2$ in $\C^2$, which 
is $w_1=\pm \ii w_0^n$. These are two copies of 
$\C$ embedded as affine algebraic curves in $\C^2$, which
intersect over the origin $\{w_0=w_1=0\}$.  
Note that when $n=1$, which is the ABJM case, 
we see $w_1=\pm \ii w_0$ are two \emph{linearly} embedded copies 
of $\C$. This is indeed the standard Hanany-Witten brane configuration 
for the conifold \cite{Mukhi}. 
For $n>1$, we obtain a \emph{non-linear} version of this, where 
the NS5-branes are embedded as the curves $w_1=\pm \ii w_0^n$ 
in $\C^2$. We label the latter directions $4589$, and 
refer to $\C^2_{4589}$. The NS5-branes also sit 
at a point in the $S^1_6$ circle, where their 
distance of separation is the period of the 
$B_2$-field through the collapsed $\mathbb{CP}^1$ in 
$W^0_n$. The final Type IIB picture is described in Figure \ref{figbranesk0}.

Note we can immediately read off the matter content of the field theory from 
this picture: the brane set-up is identical, apart from the embedding of the NS5-branes in $4589$, to the $\mathcal{A}_1$ singularity. 
Thus we may read off two gauge groups, corresponding to the $N$ 
D3-branes breaking on the two NS5-branes on the $S^1_6$ circle. 
At each NS5-brane we obtain a pair of bifundamentals, 
$A_i$, $B_i$, and an adjoint 
$\Phi_1$, $\Phi_2$ for each D3-brane segment. The $\mathcal{A}_1$ theory 
also has the $\mathcal{N}=4$ cubic superpotential for these fields.
For the $\mathcal{A}_1$ theory, both branes are parallel, say 
at the origin in the $89$ plane. For the conifold theory $n=1$, 
one brane is in the $45$ plane, while the other is in the orthogonal 
$89$ plane. This corresponds to giving a mass to the adjoints, 
-$\Phi_1^2+\Phi_2^2$, as shown in \cite{Mukhi}. Integrating these out, one obtains the quartic 
superpotential of Klebanov-Witten. In the general $n$ case, 
the non-trivial embedding of the NS5-branes in $\C^2_{4589}$ is reflected in the 
higher order $(-1)^n\Phi_1^{n+1}+\Phi_2^{n+1}$ superpotential term. 

\subsection{Adding RR-flux/D5-branes: $k\neq0$}\label{secRRflux}

The next step is to turn back on the RR two-form flux, so that $k\neq0$: this 
is then the Type IIA dual of M-theory on $X_n/\Z_k$ with $N$ spacefilling 
M2-branes.
As we discussed in section \ref{IIAsec}, supersymmetry 
also requires that one fibre the parameter $\zeta$ over the $\R_3$ 
direction. Thus, before discussing this, we first consider the 
effect of turning on the 
parameter $\zeta$ in the T-dual IIB brane set-up above. 

Without loss of generality, we take $\zeta>0$ so that 
$W^{\zeta}_n\cong W_n^+$ is biholomorphic to Laufer's 
resolved manifold, with an exceptional $\mathbb{CP}^1$ replacing the 
singular point of $W^0_n$. 
The $U(1)_6$ action on $W^0_n$ extends to an action on $W_n^+$. To see this, recall from the last part of section \ref{parent} that 
$(A_1,A_2,B_1,B_2,z_0)$ are coordinates on 
$\C^5$, and that $x=A_2B_2$, $y=A_1B_1$, $u=A_1B_2$,
$v=A_2B_1$ are invariants under $U(1)_b$, 
with $\xi=A_2/A_1$ an invariant on $U_1$ and 
$\mu=A_1/A_2$ an invariant on $U_2$. 
The embedding equation (\ref{hypersurface}) then becomes 
$x+y+z_0^n=0$. When $\zeta=0$ we have the conifold 
$xy=uv$, and eliminating $x$ this becomes 
$y^2+yz_0^n+uv=0$, which is the equation $w_1^2+w_0^{2n}=uv$ of the 
three-fold $W^0_n$ on identifying
$\ii w_1=y+\tfrac{1}{2}z_0^n$, $w_0=2^{-1/n}z_0$, as before.
Thus $U(1)_6$ rotates $u$ with charge
1 and $v$ with charge $-1$, and we may lift this 
to an action on $\C^5$ with coordinates 
$(A_1,A_2,B_1,B_2,z_0)$ by assigning charges
$(1,0,-1,0,0)$. 
It follows that the charges of 
$(x,y,u,v,\xi,\mu)$ under $U(1)_6$ are 
$(0,0,1,-1,-1,1)$. The fixed locus 
is thus $u=v=\xi=0$ and $u=v=\mu=0$ -- recall that $\xi=1/\mu$ on the 
overlap. Thus on the exceptional $\mathbb{CP}^1$ we fix 
the north pole $\xi=0$, and also the south pole 
$\mu=0$. We thus see that after resolving $W^0_n$ to $W_n^+$ the fixed 
point set under $U(1)_6$ is two disjoint copies of $\C$, over the two 
poles of the $\mathbb{CP}^1$. Indeed, recall that 
$x=-v\mu-Z_2^n$ on the patch $H_2$ (where $Z_2=z_0$), 
and thus the fixed locus at $v=\mu=0$ is described by the 
equation $x=-z_0^n$. Changing variables as above, this becomes 
precisely $w_1=-\ii w_0^n$. Conversely, 
the fixed locus $u=\xi=0$ is the equation $y=-z_0^n$, 
which under the above change of variable becomes 
precisely $w_1=\ii w_0^n$. 

One can also interpret this in the moment map picture.
The moment map is $\mu_6=
|A_1|^2-|B_1|^2$. Turning on $\zeta$, 
we also have (\ref{Dterm}). 
The exceptional $\mathbb{CP}^1$ is, 
for $\zeta>0$, at $B_1=B_2=0$. Then the moment map 
restricted to $\mathbb{CP}^1$ becomes simply 
$\mu_6\mid_{\mathbb{CP}^1} = |A_1|^2$. 
But also $|A_1|^2=\zeta-|A_2|^2$ on this locus, 
and thus we see that on $\mathbb{CP}^1$ the 
moment map ranges from $\mu_6=0$ at 
$A_1=0$ to $\mu_6=\zeta$ at $A_2=0$. 
These are precisely the two poles 
of the $\mathbb{CP}^1$, which is where the fixed 
locus is.  We thus see that the $\mathbb{CP}^1$ 
is mapped to an \emph{interval} in the image of the 
moment map $\mu_6$, which recall is the 
$\R_7$ direction, with the endpoints of the interval
being where the NS5-branes are after performing 
the T-duality along $U(1)_6$. Notice that in the holomorphic picture 
$A_1=0$ is the south pole $\mu=0$ while $A_2=0$ is 
the north pole $\xi=0$. For negative parameter $\zeta<0$, 
the roles of $A_i$ and $B_i$ swap. In this case
we will have coordinates 
$\tilde{\xi}=B_2/B_1$ and $\tilde{\mu}=B_1/B_2$ 
on the exceptional $\mathbb{CP}^1$, which is
now located at $A_1=A_2=0$. The moment map is 
$\mu_6\mid_{\widetilde{\mathbb{CP}}^1}=-|B_1|^2$. 
This ranges from $0$ at $B_1=0$ to 
$-\zeta$ at $B_2=0$, with the two endpoints 
being the NS5-brane loci. Notice that 
the brane at $-\zeta$ is $B_2=0$, which is 
$\tilde{\xi}=0$, which is the same NS5-brane 
that moves for $\zeta>0$, namely that with $w_1=\ii w_0^n$. 

To conclude, we see that the 
T-dual of resolving $W^0_n$ to $W^{\zeta}_n$ is simply 
to separate the two NS5-branes in the $\R_7$ direction 
by a distance $\zeta$ -- they are wrapped on the same 
curves as before in the $\C^2_{4589}$ direction. 
In terms of Figure \ref{figbranesk0}, the NS5-brane
on the left hand side moves a distance $\zeta$ in the 
(transverse, as drawn) $\R_7$ direction.
Notice that once we resolve $W^0_n$ there is no canonical 
place to put the D3-branes -- we have to pick a point 
on $W^{\zeta}_n$. It is natural (in the sense that it preserves a $U(1)\subset SU(2)_r$ symmetry) to put them either at the north 
pole or south pole of the $\mathbb{CP}^1$, in which case 
the D3-branes intersect either one NS5-brane or the other. 

\begin{figure}[ht!]
\epsfxsize = 15cm
\centerline{\epsfbox{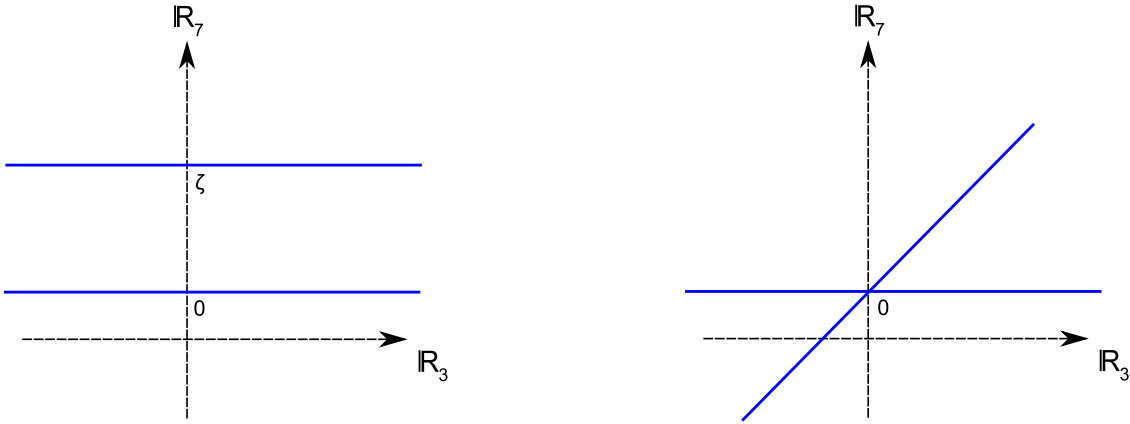}}
\caption{On the left hand side: the positions of the two NS5-branes with resolution 
parameter $\zeta$ in the Type IIA dual. The NS5-brane at position 
$\zeta$ is that wrapped on $w_1=\ii w_0^n$, while the brane at the origin 
is that wrapped on $w_1=-\ii w_0^n$.
On the right hand side: the positions 
of the 5-branes after turning on the RR flux in the Type IIA dual, which fibres 
the resolution parameter over the $\R_3$ direction. One of the branes rotates 
so that they now intersect at the origin of the $\R_3-\R_7$ plane.}
\label{figrotate}
\end{figure}

We may now consider what happens when we turn on 
the RR two-form flux. Recall this fibres the  parameter $\zeta$ over 
the $\R_3$ direction in Type IIA. It is simple to see 
what this does in the IIB brane picture. 
Consider a fixed point in $\R_3$, which means 
fixing a particular value for $\zeta$. Then the 5-branes 
are separated by some distance $\zeta$ in the $\R_7$ direction. 
More precisely, the above analysis shows that 
for $\zeta>0$ the 5-brane at the south pole is 
always at the origin in $\R_7$, while the brane 
at the north pole is at $\zeta$ in $\R_7$. 
As we move towards the origin in $\R_3$, the 5-branes 
get closer together in the $\R_7$ direction, until 
finally at the origin they meet. We may then pass 
through the origin to $\zeta<0$, where the behaviour 
is the same (with $A_i$ replaced by $B_i$). 
This shows that after turning on the RR two-form flux, 
the 5-branes rotate from being at 
fixed parallel distance in the $\R_7$ direction (and 
filling the $\R_3$ direction), to being 
two lines in the $\R_3-\R_7$ plane that cross 
at the origin -- see Figure \ref{figrotate}. This means that, after turning on the 
RR two-form flux, the 5-branes meet precisely at the 
origin in $\R^6_{345789}$. 
although they are still non-trivially holomorphically 
embedded in $\C^2_{4589}$ as 
$w_1=\pm\ii w_0^n$. 

Notice that for $n=1$ the above indeed reproduces the 
Type IIB brane picture in ABJM \cite{Aharony:2008ug} -- up to two important details. 
First, in the case $n=1$ we have derived the Type IIB 
brane dual by starting with $\C^4/\Z_k$, reducing to Type IIA 
along $U(1)_b$ and then T-dualizing to Type IIB along 
$U(1)_6$. In \cite{Aharony:2008ug}, the authors instead 
began with the Type IIB brane picture, and argued that 
T-dualizing to Type IIA and uplifting to M-theory gave 
a non-trivial hyperk\"ahler eight-manifold as the uplift, which 
is characterized by two harmonic functions, defined 
on two copies of $\R^3$. 
The difference between these two pictures is that 
the former is simply the \emph{near-brane} limit of the latter. 
Indeed, ABJM showed explicitly that the near-horizon limit 
of the hyperk\"ahler manifold indeed gives 
 $\C^4/\Z_k$, which amounts to dropping the non-zero constant 
term in the harmonic functions. This is the dual geometry in the 
region near to where the 5-branes intersect at the origin 
in $\R^6_{345789}$ (which are the two copies of $\R^3$ mentioned above). 

Second, and more importantly, in the ABJM brane picture the rotated 5-brane in 
Figure \ref{figrotate} is in fact a \emph{bound state} of an NS5-brane with $k$ D5-branes -- 
the latter is effectively the T-dual of the $k$ units (\ref{kflux}) of RR two-form flux through the (fibred) exceptional
$\mathbb{CP}^1$ in the Type IIA geometry. To see the presence of the $k$ D5-branes in the $(1,k)$5-brane 
bound state directly is not straightforward in the discussion we have given above. However, 
the $k$ units of D5-brane charge can be seen indirectly by considering a certain tadpole. 
Thus, we begin in Type IIA on $C(M_6)$, which recall may also be thought of 
as $W^{\zeta}_n$ fibred over $\R_3$. Pick a non-zero point in $\R_3$, and consider 
the exceptional $\mathbb{CP}^1$ of size $|\zeta|$ in $W^{\zeta}_n$ over this point. 
If we wrap a D2-brane over this $\mathbb{CP}^1$, we get a point particle 
in $\R^{1,2}_{012}$. However, because of the $k$ units of RR two-form flux (\ref{kflux}) 
through this $\mathbb{CP}^1$, in fact this configuration does not exist in isolation: 
one must have $k$ fundamental strings \emph{ending} on the wrapped D2-brane. 
To see this, note the Wess-Zumino coupling on the D2-brane:
\bea
\frac{1}{(2\pi)^2l_s^3} \int_{\R_{\mathrm{time}}} A \int_{\mathbb{CP}^1} F_2 
= k\cdot \frac{g_s}{2\pi l_s^2}\int_{\R_{\mathrm{time}}} A~.
\eea
To cancel this tadpole, we precisely require $k$ fundamental strings to 
end at a point on the $\mathbb{CP}^1$. 

\begin{figure}[ht!]
\epsfxsize = 15cm
\centerline{\epsfbox{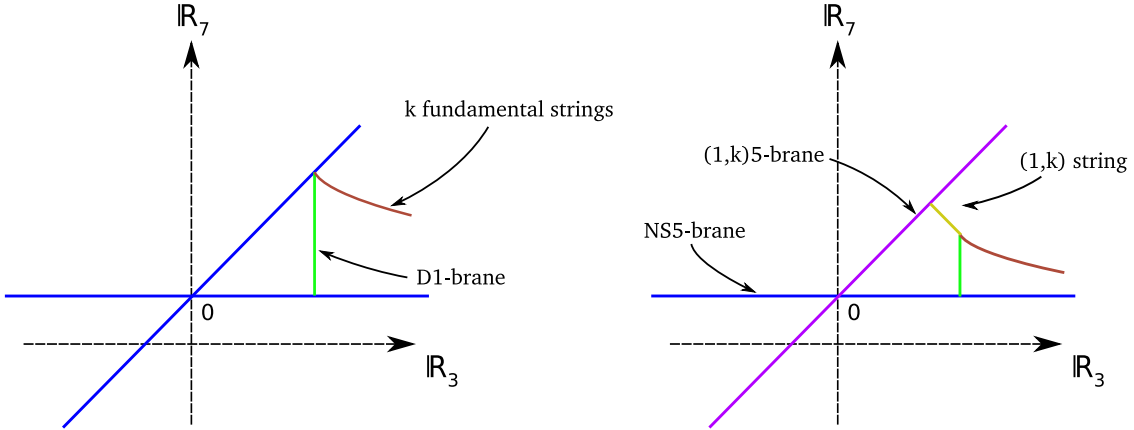}}
\caption{On the left hand side: the naive T-dual configuration to a D2-brane wrapped 
on the $\mathbb{CP}^1$ at a fixed non-zero point in $\R_3$ is a D1-brane stretching between the two NS5-branes, 
with $k$ fundamental strings also ending on the D1-brane and one of the NS5-branes to cancel the tadpole.
On the right hand side: the correct T-dual configuration, in which the D1-brane and $k$ 
fundamental strings form a $(1,k)$ string bound state, which then must necessarily end on a 
$(1,k)$5-brane. (Notice that the D1-brane must also wind around the $S^1_6$ circle as one moves 
from one 5-brane to the other along its worldvolume.)}
\label{fiD1pic}
\end{figure}

Consider the T-dual to this in Type IIB. As already discussed, 
the exceptional $\mathbb{CP}^1$ maps to an interval 
in the $\R_7$ direction, between the two 5-branes: this lies at the chosen point in 
$\R_3$, and is at the origin in $\C^2_{4589}$. A D2-brane 
wrapped on the $\mathbb{CP}^1$ thus T-dualizes to a 
D1-brane stretched between the two 5-branes in the 
$\R_7$ direction. The $k$ fundamental strings ending on the D2-brane 
T-dualize to $k$ fundamental strings ending on the D1-brane. 
In particular, the fundamental strings may end at one of the poles of the 
$\mathbb{CP}^1$. In the IIB picture, we therefore have a D1-brane 
and also $k$ fundamental strings terminating on \emph{one} of the 5-branes (while 
for the other 5-brane there is only a D1-brane ending on it). 
In general, a $(p,q)$ string, where $p$ denotes the number of D1-branes 
and $q$ the number of fundamental strings in a bound state string, can only 
end on a $(p,q)$5-brane. Thus the only way to make sense of the above 
tadpole is that the 5-brane is in fact a $(1,k)$5-brane, and the 
D1-brane and $k$ fundamental strings form a $(1,k)$ bound state ending on this. 
Of course, this precisely reproduces the correct brane configuration of 
ABJM in the case of $n=1$.

\begin{figure}[ht!]
\epsfxsize = 11cm
\centerline{\epsfbox{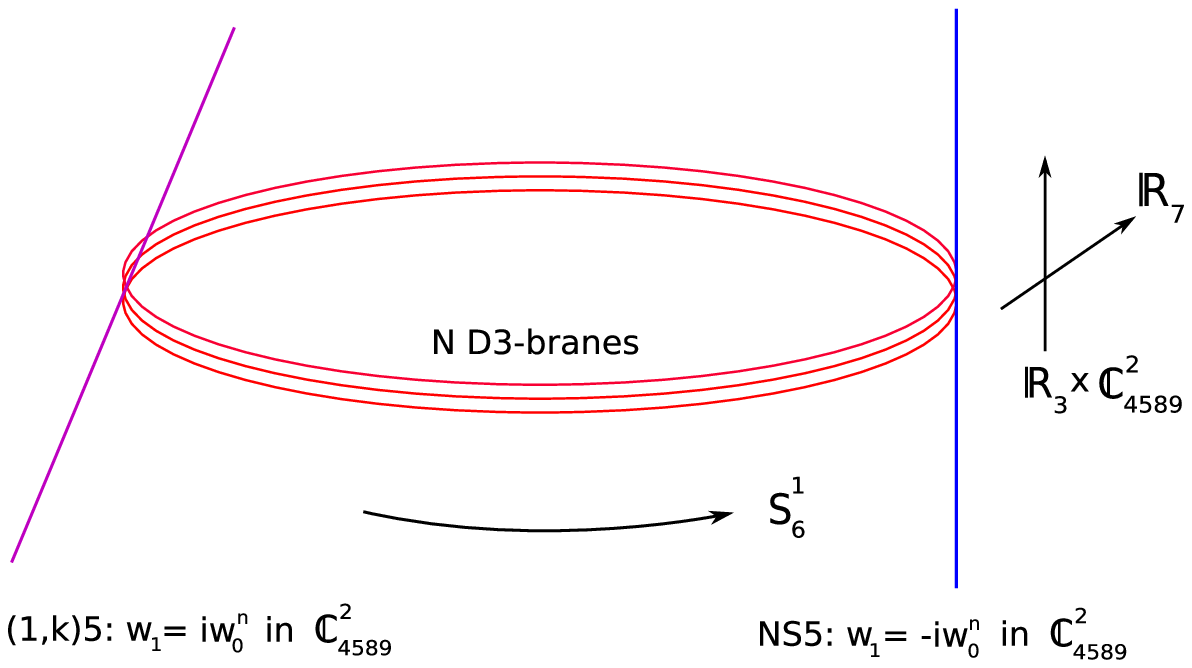}}
\caption{The final Type IIB dual of M-theory on $X_n/\Z_k$. The spacetime is 
$\R^{1,2}_{012}\times\R_3\times S^1_6\times\R_7\times\C^2_{4589}$. 
There are $N$ D3-branes filling the $\R^{1,2}_{012}$ directions and 
wrapping the $S^1_6$ circle; they are at the origin in $\R_3$, $\R_7$ and $\C^2_{4589}$. 
There are also two spacefilling 5-branes in $\R^{1,2}_{012}$ at points on the 
$S^1_6$ circle.
The first is an NS5-brane, sitting at the origin 
in $\R_7$ and filling $\R_3$, which wraps the curve 
$w_1=-\ii w_0^n$ in $\C^2_{4589}$. The second is 
a $(1,k)$5-brane, wrapping 
an angled line through the origin in the $\R_3-\R_7$ plane, 
and wrapping the curve $w_1=\ii w_0^n$ in $\C^2_{4589}$.}
\label{figbranepicture}
\end{figure}

To conclude, we have shown that M-theory on $X_n/\Z_k$ has a Type IIB dual of 
Hanany-Witten type: it is identical to the brane set-up for $n=1$ described 
by ABJM \cite{Aharony:2008ug}, except that the 5-branes are wrapped on 
the holomorphic curves $w_1=\pm \ii w_0^n$ inside $\C^4_{4589}$ -- see
Figure \ref{figbranepicture}. 

\subsection{Brane creation effect}

Having described the Type IIB brane dual, an important dynamical question 
is what happens when we move the two 5-branes past each other on the 
$S^1_6$ circle. This was first studied by Hanany-Witten \cite{Hanany:1996ie}, 
and the analysis in section 5 of that paper may be applied directly 
to the case $n=1$ (the ABJM case). We thus begin by describing 
the $n=1$ case, and then explain how to apply this result for 
$n>1$ by deforming the curves in $\C^2_{4589}$ so that 
the brane intersections in $\R^6_{345789}$ are normal crossings.

We thus start with $n=1$. We suppress the spacetime $\R^{1,2}_{012}$ from 
the discussion, since all branes are spacefilling in these directions. 
Thus the relevant geometry is $S^1_6\times\R^6_{345789}$.
We have an NS5-brane at a point $0\neq t \in S^1_6$ and at the origin 
in $789$, and a $(1,k)$5-brane at the origin $0\in S^1_6$ 
and at the origin in $345$. Notice that we have, for convenience of notation, rotated the 
axes relative to Figure \ref{figbranepicture}: the argument we are about to 
give is entirely topological, and so is unaffected. We denote these submanifolds 
as $W_{NS,t}$ and $W_{(1,k)}$, respectively.
These two copies of 
$\R^3$ that are wrapped by the 5-branes thus intersect normally at the origin 
in $\R^6_{345789}$. However, importantly, the branes do not actually intersect in spacetime 
unless $t=0$. 

The $(1,k)$5-brane sources $k$ units of RR three-form flux $F_3$ through a sphere $S^3$ linking its 
worldvolume. Thus, let $S^3$ be a normal sphere around a point on the $(1,k)$-brane 
in $S^1_6\times\R^6_{34578}$, so that
\bea
\frac{1}{(2\pi l_s)^2g_s}\int_{S^3} F_3 = k~.
\eea
Following \cite{Hanany:1996ie}, we then define the \emph{linking number} 
\bea\label{link}
L_t = \frac{1}{(2\pi l_s)^2g_s}\int_{W_{NS,t}} F_3~.
\eea
This is independent of $t$ as $t$ is varied, \emph{provided} we do not cross 
the origin $t=0$. The reason for this is that $F_3$ is closed on the complement 
of the $(1,k)$5-brane worldvolume, and the independence of (\ref{link}) on $t$ 
then follows from Stokes' Theorem. More precisely, $\diff F_3$ is a 
four-form which is supported only on the $(1,k)$5-brane worldvolume at $t=0$ and 
the origin in $345$: 
it is $k$ times a delta-function representative of the Poincar\'e dual of 
$W_{(1,k)}$.

Consider now moving the NS5-brane from $t_+>0$, on the right of the $(1,k)$5-brane, 
to $t_-<0$ on the left. Let $I=[t_-,t_+]$ be the interval in the $S^1_6$ circle covered 
in this motion. Then we have linking numbers (\ref{link}) $L_+$ and $L_-$ on the right 
and left. We may compute the change in linking number using Stokes' Theorem:
\bea\label{changelink}
L_+ - L_- =  \frac{1}{(2\pi l_s)^2g_s}\int_{W_{NS}\times I} \diff F_3 = k~.
\eea

On the worldvolume of the NS5-brane there is a $U(1)$ gauge field $A_{NS}$, with field 
strength $F_{NS}$, and 
it is only the combination $\Lambda = C_2 - 2\pi l_s^2 F_{NS}$ that is gauge invariant. Moreover, 
\bea\label{tady}
F_3\mid_{W_{NS}} = \diff \Lambda~,
\eea
meaning that $F_3$ must be exact on the NS5-brane worldvolume $W_{NS,t}$. In the non-compact 
setting of interest, of course all closed forms are exact on $W_{NS,t}\cong \R^3$, so  
(\ref{tady}) is always satisfied. However, what we learn from (\ref{changelink}) is 
that the period of $F_3$ through $W_{NS,t}$ changes by $k$ units as we move the NS5-brane 
from the right $t>0$ to the left $t<0$ of the $(1,k)$5-brane. The explanation for this is that 
$k$ spacefilling D3-branes are created at the intersection point $t=0$ when the branes 
are moved past each other. Indeed, such a D3-brane ending on the NS5-brane is a delta-function 
source for $F_{NS}$:
\bea\label{D3tad}
\frac{1}{2\pi g_s}\diff F_{NS} = \pm \delta(p)
\eea
where $p\in W_{NS}\cong \R^3$. That is, the D3-brane ending on the NS5-brane is a 
magnetic monopole for this $U(1)$ gauge field.
The sign in (\ref{D3tad}) depends on whether the D3-brane ends from the right or 
from the left on the $S^1_6$ circle, which it wraps (a monopole or anti-monopole). Integrating $k$ times (\ref{D3tad}) over $W_{NS}$ precisely 
accounts for the change in linking number (\ref{changelink}). This is the Hanany-Witten effect.

Having carefully reviewed this effect, we may now apply it to the case with $n>1$. 
However, note that for $n>1$ the branes are not linearly embedded in $\C^2_{4589}$: they 
cross at a single point at the origin, but they are wrapped on non-trivial curves. 
We may remedy this by \emph{deforming} the curves that the 5-branes are wrapped on. Thus, 
we change 
\bea\label{def1}
w_1=-\ii w_0^n &\longrightarrow& w_1 = -\ii\prod_{i=1}^n (w_0-\alpha_a) + \alpha_0\\\label{def2}
w_1=\ii w_0^n &\longrightarrow& w_1 = \ii\prod_{i=1}^n (w_0-\beta_a) + \beta_0~.
\eea
Here $\alpha_a$, $\beta_a$, $a=0,\ldots,n$, are arbitrary parameters. The point of these deformations 
is that $(a)$ they preserve the boundary conditions at infinity, since we have added 
only lower order terms to the polynomials, and $(b)$ the resulting curves now intersect 
normally in $\C^2_{4589}$. Indeed, these two curves in $\C^2_{4589}$ intersect where 
the $w_1$ coordinate in (\ref{def1}) equals the $w_1$ coordinate in (\ref{def2}). This 
results in the $n$th order polynomial
\bea
\ii\prod_{i=1}^n (w_0-\alpha_a)+ \ii\prod_{i=1}^n (w_0-\beta_a) - \alpha_0 +\beta_0=0~.
\eea
For \emph{generic} values of the parameters $\alpha_a$, $\beta_b$, this will have 
precisely $n$ solutions for $w_0$, say $w_0^{(i)}$, $i=1,\ldots,n$. Thus 
the resulting curves generically intersect at $n$ points $(w_0^{(i)},w_1^{(i)})$,
where of course $w_1^{(i)}$ is given by (\ref{def1}) (or (\ref{def2})) evaluated 
at $w_0^{(i)}$. Moreover, the intersects of the curves near to these $n$ points 
look precisely like the linear $n=1$ case. 

We are now in good shape: after this generic deformation that preserves the boundary conditions 
of the branes at infinity, the two branes intersect ordinarily at $n$ points in 
$\R^6_{345789}$ (they always cross  at the origin of the $\R_3-\R_7$ plane). 
The above discussion of the Hanany-Witten effect shows that the creation 
of the $k$ D3-branes as an NS5-brane crosses a $(1,k)$5-brane occurs entirely 
\emph{locally} at the points where the branes intersect in spacetime. 
Thus if we move our deformed NS5-brane past the deformed $(1,k)$5-brane, 
we obtain precisely $n$ copies of the $n=1$ result, {\it i.e.} in total 
$nk$ D3-branes are created as they are moved past each other. 
More precisely, $k$ D3-branes are created at each of the $n$ points 
$(w_0^{(i)},w_1^{(i)})$ (at the origin in the $\R_3-\R_7$ plane, and stretched
along the $S^1_6$ circle). Notice that this result is independent of 
the choice of deformation parameters $\alpha_a$, $\beta_a$, as it is topological. 
Thus after moving the branes past each other we may deform back to 
$\alpha_a=\beta_a=0$, where the $nk$ created D3-branes are all at the origin
in $\R^6_{345789}$. 

\subsection{The field theory duality}
\label{ftd}

\begin{figure}[ht!]
  \begin{minipage}[t]{0.48\textwidth}
    \begin{center}
    \epsfxsize=7cm
    \epsfbox{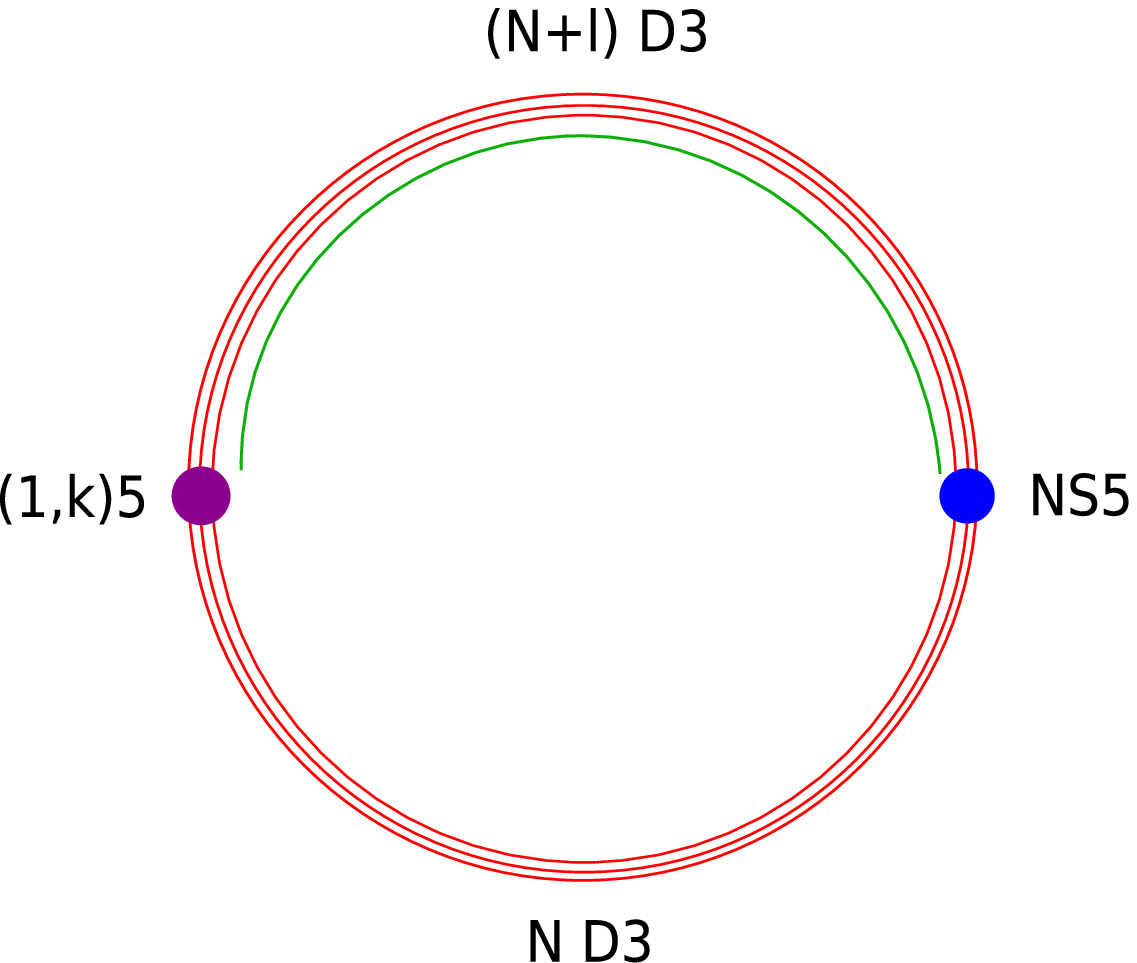}
      \end{center}
  \end{minipage}
  \hfill
  \begin{minipage}[t]{.48\textwidth}
    \begin{center}
     \epsfxsize=5cm
    \epsfbox{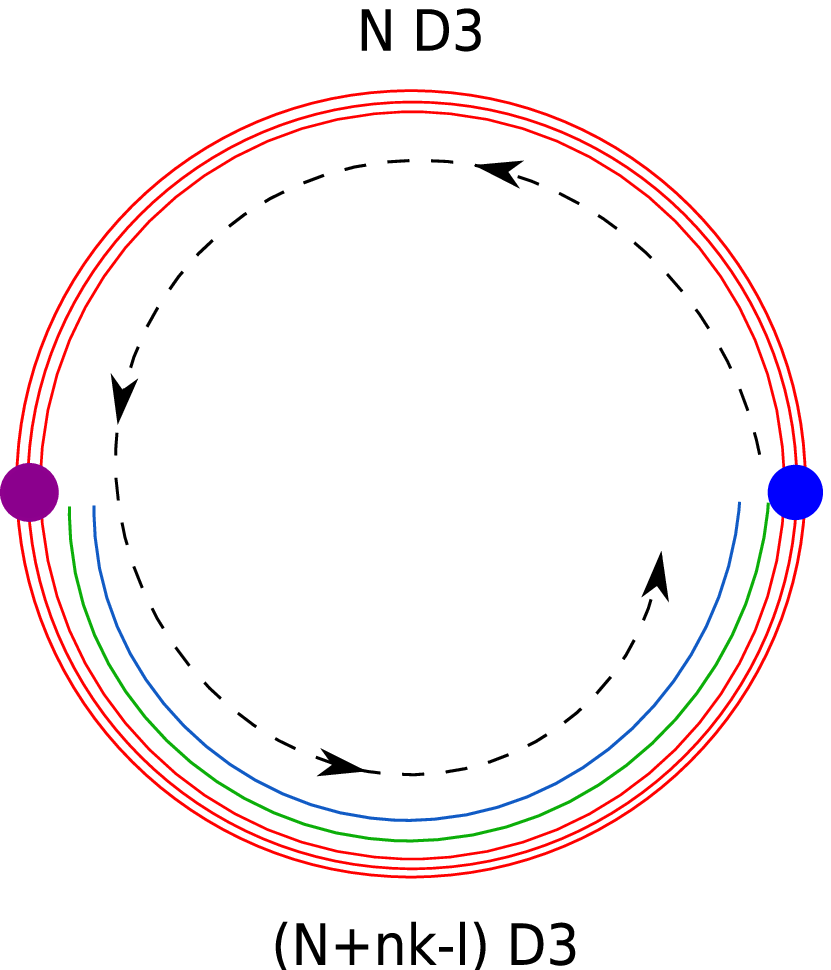}
     \end{center}
  \end{minipage}
\caption{On the left hand side: the initial brane configuration, with $(N+l)$ D3-branes suspended between the 
5-branes on one side of the $S^1_6$ circle, and $N$ D3-branes on the other. On the right hand side: 
moving the NS5-brane anti-clockwise around the circle pulls the $l$ fractional 
branes with it. After passing the $(1,k)$5-brane these swap orientation, becoming $l$ anti-branes, and in addition $nk$ D3-branes 
are created. }
  \hfill
\label{rotatingfig}
\end{figure}

The brane creation effect described in the last section leads to an interesting field theory duality, discussed 
for the ABJM theory in \cite{Aharony:2008gk}, \cite{Aharony:2009fc}. Here we briefly describe the 
situation for general $n$.
We begin with the Type IIB brane set-up corresponding to the gauge group $U(N+l)_k\times U(N)_{-k}$. This is shown 
on the left hand side of Figure \ref{rotatingfig}.

Consider, without loss of generality, moving the NS5-brane around the circle. Rotating it anti-clockwise by one 
revolution, as shown on the right hand side of Figure \ref{rotatingfig}, 
the gauge groups become $U(N)_k\times U(N+nk-l)_{-k}$.
In particular, we note 
that the $U(N+nk)_k\times U(N)_{-k}$ theory can be deformed to the 
$U(N)_k\times U(N)_{-k}$ theory in this way, which is the required field theory duality 
to match the dual supergravity analysis mentioned at the very end of section \ref{vmsection}. Moving the NS5-brane multiple times around the circle, or in the other direction, apparently leads to further 
equivalences, as observed for the $n=1$ ABJM theory in \cite{Aharony:2008gk}. This certainly deserves further careful study 
of the brane system to understand properly, although we shall make some comments on this in section \ref{sec:IR}.


\section{The deformed supergravity solution}
\label{defsol}

In this section we describe a supergravity solution \cite{Cvetic:2000db} which is a deformation of the AdS$_4\times V_{5,2}/\Z_k$
M-theory background discussed in section \ref{mdual}, in the sense that it approaches 
the latter asymptotically at infinity. Throughout this section we set $n=2$. We also begin with $k=1$, and restore general $k$ later.

\subsection{The Stenzel metric on $T^*S^4$}

We begin by describing a deformation of the Calabi-Yau cone metric on
the quadric cone $X_2$. The latter has an isolated singularity at 
$z_0=\cdots=z_4=0$ that may be \emph{deformed}\footnote{In the same sense as the more familiar deformed conifold in six dimensions.} to a smooth non-compact Calabi-Yau variety $\cX$,  
diffeomorphic to $T^*S^4$ (the cotangent bundle of $S^4$), via 
\bea\label{deformation}
\cX\equiv \left\{\sum_{i=0}^4 z_i^2 = \gamma^2\right\}~,
\eea 
where $\gamma\in \C$ is a constant. For $\gamma\neq 0$ this describes a smooth complex 
structure on $T^*S^4$. The deformation breaks the $\C^*\cong\R_+\times U(1)_R$ symmetry of the cone 
to $\Z_2\subset U(1)_R$. Using the broken $U(1)_R$ action we take $\gamma\in\R_+$ in what follows.
The $S^4=SO(5)/SO(4)$ zero-section is then realized as the real locus of $\cX$ in $\C^5$. The cotangent bundle structure may be seen explicitly  by writing
\bea
z_i = \cosh \left(\sqrt{p_jp_j}\right)x_i + \frac{\mathrm{i}}{\sqrt{p_jp_j}}
\sinh \left(\sqrt{p_jp_j}\right) p_i~.
\eea
Then $\sum_{i=0}^4 x_i^2 = \gamma^2$, $\sum_{i=0}^4 x_ip_i=0$, so that  the $S^4$ is $\{p_i=0\}$.

There is an explicit complete Ricci-flat K\"ahler metric on $\cX$
which is asymptotic to the cone metric at large radius, called the 
Stenzel metric. This is cohomogeneity one under the action of $SO(5)$, 
with principal orbits diffeomorphic to $V_{5,2}=SO(5)/SO(3)$, and 
degenerate special orbit $S^4=SO(5)/SO(4)$.
The K\"ahler structure induces the standard symplectic structure on $T^*S^4$, 
and thus the $S^4$ is Lagrangian; in fact it is special Lagrangian, and is thus 
a minimal volume representative of the generator of $H_4(\cX,\Z)\cong\Z$. 
Note that given any Ricci-flat metric $\diff s^2$, the rescaled metric 
$\gamma^2 \diff s^2$ is also Ricci-flat, for any positive constant $\gamma\in\R_+$, 
and this is essentially the constant $\gamma$ above, which is proportional to 
the radius of the $S^4$. 

In terms of invariant one-forms on the coset space $V_{5,2}=SO(5)/SO(3)$,
the metric on $\cX$ may be written as 
\bea
\diff s^2_{\cX} \,=\,  c^2 \diff r^2 + c^2 \nu^2 + a^2\sum_{i=1}^3 \sigma_i^2 + b^2 \sum_{i=1}^3\tilde{\sigma}_i^2~,
\eea
where 
\bea
&&a^2 = \frac{1}{3}(2+\cosh 2r)^{1/4}\cosh r~, \qquad b^2 = \frac{1}{3}(2+\cosh 2r)^{1/4}\sinh r\tanh r~,\nonumber\\
&& c^2 = (2+\cosh 2r)^{-3/4} \cosh^3r~.
\label{stenzsol}
\eea
More details may be found in appendix \ref{appdetails}.
In these coordinates, the $S^4$ is located at $r=0$. Note here we have picked a particular 
representative metric in the conformal class of metrics on $\cX$, {\it i.e.} a particular value of $\gamma$. 
It will be straightforward to reintroduce this scale later. The calibrated $S^4$ in the above 
solution has fixed size, with induced round metric
\bea
\diff s^2_{S^4} = 3^{-3/4}(\nu^2 +\sum_{i=1}^3\sigma_i^2)~.
\eea 
After a change of variable
\bea
\rho^2 \sim \frac{16}{9} \frac{1}{2^{9/4}}\mathrm{e}^{\frac{3}{2}r}~,
\label{rchange}
\eea
the asymptotic form of the metric is 
\bea
\diff s^2 \approx  \diff \rho^2 + \rho^2 \left[\frac{3}{8}\sum_{i=1}^3\left( \sigma_i^2 + \tilde{\sigma}_i^2\right)  +\frac{9}{16} \nu^2  + \frac{2^{1/3}}{3^3} \frac{1}{\rho^{8/3}}\sum_{i=1}^3\left( \sigma_i^2 - \tilde{\sigma}_i^2\right) + \dots\right] ~.
\eea
The leading term is the metric on the cone over the manifold $Y_2=V_{5,2}$.

For later use we record here the results of certain integrals.  Noticing that the $S^4$ is parametrized by $\nu, \sigma_i$,
and recalling that  $V_{5,2}$ is an $S^3$ bundle over $S^4$, we have
\bea
\int_{S^3_{\mathrm{fibre}}} \tilde \sigma_1 \wedge \tilde \sigma_2 \wedge \tilde \sigma_3 = 2\pi^2~.
\eea
This is the volume of a unit $S^3$, as necessarily follows since the collapse of this $S^3$ 
at the $S^4$ zero-section is regular. Writing the volume form of $V_{5,2}$ as
\bea
\diff \mathrm{vol}_{V_{5,2}} \,=\, \frac{3^4}{2^{11}} \sigma_1 \wedge \sigma_2 \wedge \sigma_3\wedge  \tilde \sigma_1 \wedge \tilde \sigma_2 \wedge \tilde \sigma_3 \wedge \nu~,
\eea
and using the total volume  of $V_{5,2}$ (\ref{totvol}), we deduce also that  
\bea
\int_{S^4}\nu\wedge\sigma_1 \wedge \sigma_1 \wedge \sigma_3 = \frac{8\pi^2}{3}~,
\eea
which is in fact the volume of a unit radius round $S^4$.

\subsection{The deformed M2-brane solution}

The AdS$_4\times V_{5,2}$ supergravity solution admits 
a smooth supersymmetric  deformation, based on the above Stenzel metric. 
This solution was presented in \cite{Cvetic:2000db}.  
We have found and corrected a few minor mistakes in the 
formulas in \cite{Cvetic:2000db}, which are
important for the physical interpretation. The $d=11$ solution is\footnote{We have introduced an explicit deformation parameter 
$\gamma$ which is set to unity in  \cite{Cvetic:2000db}. This measures the radius of the $S^4$ at the origin.}  
\bea\label{solnn}
\diff s^2 &=& H^{-2/3}\diff s^2_{\R^{1,2}} + H^{1/3} \gamma^2 \diff s^2_\cX~,\nonumber\\
G &=& \diff^3 x\wedge \diff H^{-1} + m\alpha~,
\eea
where $m$ is a constant, $\diff s^2_\cX$ denotes the Stenzel metric, and $\alpha$ is a harmonic 
\emph{self-dual} four-form on $\cX$ \cite{Cvetic:2000db}. 
In terms of the orthonormal frame (\ref{frame}) defined 
 in appendix \ref{appdetails} this reads 
\bea\label{alpha}
\alpha \,=\, \frac{3}{\cosh^4r} \left(e^{\tilde{0}123}+e^{0\tilde{1}\tilde{2}\tilde{3}}\right) 
+ \frac{1}{2}\frac{1}{\cosh^4 r}\epsilon_{ijk} \left(e^{0ij\tilde{k}}+e^{\tilde{0}i\tilde{j}\tilde{k}}\right)~.
\eea
More precisely, this is an $L^2$-normalizable primitive harmonic $(2,2)$-form on $\cX$. 
Note that $\alpha$ generates $H^4_{\mathrm{cpt}}(\cX,\R)\cong \R$. By the general results of \cite{Hausel}, 
this is the only $L^2$-normalizable harmonic form on $\cX$ in fact. The equation of motion for the $G$-field 
\bea\label{fourformeqn}
\diff * G \,= \,\frac{1}{2}G\wedge G~,
\eea
implies the following equation for  the warp factor 
\bea
\Delta_\cX H \,=\, -\frac{12m^2}{\cosh^8 r}~.
\label{warpeq}
\eea
Here $\Delta_\cX$ denotes the scalar Laplacian on the Stenzel manifold with metric $ \diff s^2_\cX$.
This can be integrated explicitly in terms of the variable  $y^4 = 2 + \cosh 2r$, giving 
\bea
H(y) \,=\, \frac{-24m^2}{\sqrt{2}}\int \frac{\diff y}{(y^4-1)^{5/2}}~,
\eea
where an integration constant has been fixed by requiring regularity near to $r=0$.
In terms of the variable $\rho$ introduced in (\ref{rchange}), the asymptotic expansion reads
\bea
H(\rho)  \,=\, \frac{2^{10}}{3^5}\frac{m^2}{\rho^6} + \dots  \qquad \mathrm{for}~ \rho \to \infty~.
\label{warpy}
\eea
Notice that this has a different behaviour from the Klebanov-Strassler solution, where one has logarithmic corrections. As explained in \cite{Cvetic:2000db}, this difference comes from the fact that the self-dual harmonic form is \emph{normalizable} here, while it is not normalizable in six dimensions.  At large $\rho$ the solution becomes of the form (\ref{adssolution}), where here 
the AdS$_4$ radius is expressed in terms of the integration constant $m^2$ as 
$R^6 \,=\, \tfrac{2^{10}}{3^7}m^2$.

\subsection{The $G$-flux}

We now wish to discuss the quantization of the flux, thus relating the constant $m^2$ to the quantized fluxes. 
Because the background is asymptotically AdS$_4\times V_{5,2}$, it is natural to quantize the flux of $*G$
through the $V_{5,2}$ at infinity, as in (\ref{gquant}), and interpret this as the number of M2-branes in the UV.
More generally, we may define a ``running'' number of M2-branes $N(r)$ as
\bea
N(r) \,=\, \frac{1}{(2\pi l_p)^6}\int_{Y_r} *  G~,
\eea
where the integral is evaluated on a seven-dimensional surface of constant $r$, which is a copy of $V_{5,2}$.
To compute this, we may use the four-form equation of motion (\ref{fourformeqn})
to write
\bea
\int_{Y_2^r} *  G = \frac{1}{2}\int_{\cX^r} G \wedge G = \frac{1}{2} \int_{\cX^r} m^2 |\alpha|^2 \diff \mathrm{vol}_\cX~,
\eea
where the integral is evaluated on the Calabi-Yau $\cX$ cut off at a distance $r$. The result is 
\bea
N(r) =\frac{1}{(2\pi l_p)^6} \frac{m^2}{9} \frac{2^{11}}{3^4}\mathrm{vol}(V_{5,2}) \tanh^4 r~.
\eea
We see that this running number of M2-branes becomes a constant at infinity, where
\bea
N \,\equiv \, N(\infty) \,= \, \frac{1}{(2\pi l_p)^6} \frac{2^{11}}{3^6} m^2 \mathrm{vol}(V_{5,2})~.
\eea
This determines $m^2$ in terms of the physical paramater $N$.
Eliminating $m^2$ we see that the (UV) AdS$_4$ radius takes  exactly  the form (\ref{bigradius}).

We are not quite done, however. There is a non-trivial cycle in the geometry, namely the 
four-sphere at the zero-section of $\cX=T^* S^4$. Thus we have to impose the quantization of the four-form flux through this 
cycle. Noting that the restriction of the $(2,2)$ four-form $\alpha $ to a four-sphere at any distance $r$ 
from the origin is 
\bea
\alpha|_{S^4_r} = \frac{1}{\sqrt{3}\cosh r} \nu \wedge \sigma_1 \wedge \sigma_2 \wedge \sigma_3~,
\label{hyppo}
\eea
we compute 
\bea
\frac{1}{(2\pi l_p)^3}\int_{S^4} G \,=\, \frac{1}{(2\pi l_p)^3} \frac{m}{\sqrt{3}} \frac{8\pi^2}{3}\,= \,
\tilde M \in \mathbb{N}~,
\label{quantm}
\eea
where recall that the volume of the unit $S^4$ at the origin is $8\pi^2/3$. 
The reason for denoting the integer\footnote{It is again important here that the membrane anomaly on 
$\cX$ vanishes. This follows from the fact that $w_4(\cX)\mid_{S^4}$ is twice the fourth Stiefel-Whitney class of 
the bundle $TS^4$, and hence zero mod 2 (the latter Stiefel-Whitney class also happens to be zero). } flux as $\tilde M$ will become clear momentarily.
We hence obtain another expression for $m^2$, namely $m^2=27 \pi^2 l_p^6  \tilde M^2$.
The running number of M2-branes then takes the simple form
\bea\label{running}
N(r)\, = \, \frac{\tilde M^2}{4} \tanh^4 r~.
\eea
There is a simple way to check the numerical factor here. If we integrate (\ref{fourformeqn}) 
over the whole of $\cX$, the left hand side gives $(2\pi l_p)^6 N$.
On the other hand, the right hand side is a topological 
quantity. To see this, note that the integral of $G$ over $S^4$ is by definition $(2\pi l_p)^3 \tilde M$. 
But we may also regard $G$ as defining an element of $H^4_{\mathrm{cpt}}(\cX,\R)$. The map
$\R\cong H^4_{\mathrm{cpt}}(\cX,\R)\rightarrow H^4(\cX,\R)\cong \R$ is just multiplication by 2, the 
latter being the Euler number of $S^4$. Then we may interpret $\frac{1}{2}\int_\cX G\wedge G$ 
as the cup product $H^4(\cX,\R)\times H^4_{\mathrm{cpt}}(\cX,\R) \rightarrow H^8_{\mathrm{cpt}}(\cX,\R)=\R$ via $\frac{1}{2}[G]\cup [G]_{\mathrm{cpt}} = (2\pi l_p)^6 \frac{1}{2} \tilde M \cdot \frac{\tilde M}{2}$. 
This is a simple topological check on (\ref{running}).

Since we have $N=\tilde M^2/4$, and $N$ must be an integer, we have to set $\tilde M = 2 M$. We thus obtain the relation
\bea
N \, =\, M^2~,
\eea
where $2M$ is the number of units of $G$-flux through the $S^4$ (\ref{quantm}).
Notice that the higher derivative $X_8$ term in M-theory would lead 
to a $O(1/N)$ correction to this  formula. In fact an explicit solution, generalizing that 
above and including the $X_8$ correction, was given in \cite{CveticX8}\footnote{Although some 
errors in \cite{Cvetic:2000db} have propagated to this reference.}. Of course, the supergravity 
solution is only valid at large $N$ (and hence large $M$) 
in any case, and this term is  a subleading correction.

As a consequence of the relation $\tilde M = 2 M$ we also see that 
there is \emph{no} torsion $G$-flux turned on in $H^4(V_{5,2},\Z)\cong \Z_2$. To see this we recall
that there is a relation between the cohomology of the deformed space $\cX$ and the cohomology of its
boundary $\de \cX =V_{5,2}$. The only non-trivial cohomology of $\cX$ is 
$H^4(\cX,\Z)\cong H_4(\cX,\Z)\cong \Z$, the latter being generated 
by the $S^4$ zero-section. There is a map 
$\Z\cong H^4(\cX,\Z)\rightarrow H^4(V_{5,2},\Z)\cong\Z_2$ 
induced by restriction to $V_{5,2}=\partial \cX$ which is simply reduction 
modulo 2. 
The calculation (\ref{quantm}) means that as a cohomology class $[G] = 2 M e$, where  $e$ denotes the generator of  $H^4(\cX,\Z)$. This then maps  $[G] \to 0 \in H^4(V_{5,2},\Z) \cong H_3 (V_{5,2},\Z)\cong \Z_2$. 

We may also define a ``running $C$-field period''. 
Recall that $V_{5,2}$ may  be thought of as an $S^3$ bundle over $S^4$. Then the generator of $H_3(V_{5,2},\Z)\cong H^4(V_{5,2},\Z)\cong \Z_2$
may be taken to be a copy of the $S^3$ fibre at a fixed point on the base $S^4$. 
We can identify the torsion three-cycle at 
a distance $r$ as the three-sphere at a distance $r$  from the origin of the fibre $\R^4$, at a fixed point on $S^4$.  
We have
\bea
\alpha|_{\R^4} = \frac{\sinh^3 r }{\sqrt{3}\cosh^4 r} \diff r \wedge \tilde \sigma_1 \wedge \tilde \sigma_2\wedge \tilde \sigma_3 ~,
\eea
and thus 
\bea
c_3 (r) \,\equiv \, \frac{1}{(2\pi l_p)^3} \int_{S^3_r} C \,= \, \frac{m}{(2\pi l_p)^3} \int_{\R^4_r} \alpha \,= \,  \frac{M}{2} \left[ \frac{1}{\cosh r} \left(\frac{1}{\cosh^2 r} - 3\right)+2 \right] \label{krun}~.
\eea
Notice that  $c_3(\infty)= M$.  Indeed, this is again purely a topological integral, namely
$(1/(2\pi l_p)^3)\int_{\R^4_{\mathrm{fibre}}} G = M$, and shows that 
the holonomy of the $C$-field on $V_{5,2}$ at infinity is indeed trivial, {\it cf} 
(\ref{Cperiod}).

\subsection{The $\Z_k$ quotient}

If we wish to consider deformations of the $\vmodk$ supergravity background with $k>1$, 
the deformed solution $\cX/\Z_k$ is then singular, having two isolated $\C^4/\Z_k$ singularities at the north $p_N$ and south $p_S$ poles of the $S^4$ zero-section. 
Since we cannot trust the supergravity solution near to these points, we should remove 
them from the spacetime in any supergravity analaysis. It then makes sense to analyse flux quantization 
on the smooth manifold $(\cX\setminus \{p_N,p_S\})/\Z_k$. This has a boundary with three connected components: 
$\vmodk$ at infinity, and two copies of $S^7/\Z_k$ near to $r=0$.

Since $H_4(\cX,\Z)\cong\Z$, generated by 
the $S^4$ zero-section, it follows from a simple Mayer-Vietoris sequence that also $H_4(\cX\setminus\{p_N,p_S\},\Z)\cong\Z$. 
On removing the two points, the image of the $S^4$ zero-section in $\cX\setminus\{p_N,p_S\}$ is 
$I\times S^3$, where $I$ is an interval. Thus the image of this 
$S^4$ naturally gives a relative class in $H_4(\cX\setminus\{p_N,p_S\}, S^7\amalg S^7,\Z)$, 
although again it is simple to show that this is isomorphic to $H_4(\cX\setminus\{p_N,p_S\},\Z)$ 
and thus the relative class is represented by a closed 4-cycle also. 

Consider a $\Z_k$-invariant closed four-form $G$ on $\cX$ that has non-zero integral over the $S^4$. 
Then one obtains a four-form on $(\cX\setminus\{p_N,p_S\})/\Z_k$ with non-zero integral 
over $I\times S^3/\Z_k$, where $\Z_k$ acts along the Hopf fibre of the $S^3$. We now normalize the flux $G/(2\pi l_p)^3$ to have period 
$\tilde{M}\in\Z$ through this (relative) 4-cycle. It follows that lifting to the covering 
space $\cX$, we obtain a period $k\tilde{M}$ through $S^4$. Then the integral of $(2\pi l_p)^{-6}\tfrac{1}{2} G\wedge G$ 
over the covering spacetime $\cX$ may be carried out as in the smooth case, to give $\tfrac{1}{2}\cdot (k\tilde{M})\cdot \tfrac{1}{2}
(k\tilde{M}) = k^2 M^2$. Thus on the quotient $\cX/\Z_k$ we obtain
\bea
 N \,=\,\frac{1}{(2\pi l_p)^6} \int_{\vmodk} * G \, =\,  \frac{1}{(2\pi l_p)^6}\int_{\cX/\Z_k}\frac{1}{2}G\wedge G \,=\,  k M^2~. 
\eea
Similarly, we have
\bea
\frac{1}{(2\pi l_p)^3}\int_{\R^4_{\mathrm{fibre}}/\Z_k} G = \frac{1}{(2\pi l_p)^3}\int_{\Sigma^3} C = M~,
\eea
where we have noted that the generator $\Sigma^3$ of $H_3(\vmodk,\Z)\cong\Z_{2k}$ is given by a copy of 
the boundary of the $\R^4/\Z_k$ fibre of $T^*S^4/\Z_k$ over the north pole $p_N\in S^4$. Comparing to 
(\ref{Cperiod}), we see that $l\cong 0$ mod $2k$ at infinity, and hence there are \emph{no fractional M5-branes}.
Clearly, this is in stark contrast to the Klebanov-Strassler solution.

\section{The deformation in the field theory}
\label{defsection}

The deformed supergravity background that we have discussed is of a type which has no known 
counterpart in the context of
the AdS$_5$/CFT$_4$ correspondence. This was already noticed in 
\cite{Cvetic:2000db,Herzog:2000rz,Cvetic:2001zb}.
 The UV region is asymptotic to a Freund-Rubin background 
AdS$_4\times Y^7$, and thus according to the AdS/CFT dictionary it should be dual to the  
conformal Chern-Simons-quiver theory extensively 
discussed in the paper.  
On the other hand, in the IR region the solution is smooth and 
displays a finite-sized minimal submanifold at the bottom of the throat. 
Therefore, according to the 
general rules of gauge/gravity duality,
the dual field theory should have a mass gap and is presumably confining  \cite{Witten:1998qj}. 
Understanding the precise mechanism in the field theory is clearly an interesting challenge.  
In this final section we take a few  steps in this direction, 
leaving a more detailed investigation for future work. 

\subsection{The field theory in the UV}

As we have already explained, at infinity the deformed solution approaches the AdS$_4\times \vmodk$ background. 
Since $H^4(\vmodk,\Z)\cong\Z_{2k}$, at infinity we can only have a flat torsion $G$-flux of $[G]=l$ mod $2k$. 
A careful examination of flux quantization in the deformed solution leads to $2M$ units of $G$-flux through the 
minimal four-cycle $S^4/\Z_k$ at the zero-section $r=0$. However, this $G$-field descreases as we move towards the UV, 
eventually disappearing at infinity $r=\infty$. The topological class of this $G$-flux at infinity is $[G]=0$, while 
the flux of $*G$ through $V_{5,2}$ is $N=kM^2$. This leads us to conjecture that the field 
theory in the UV is the superconformal 
Chern-Simons-quiver theory with gauge group 
\bea
U(kM^2)_{k} \times U(kM^2)_{-k}~.
\eea 
Note that the ranks of the gauge groups could receive subleading corrections that may
 be important for a consistent
interpretation.

On general grounds, the field theoretic interpretation of the deformation
is either a perturbation by a relevant operator in the Lagrangian, or involves 
spontaneous symmetry breaking.  These two possibilities are distinguished by 
the asymptotic behaviour of perturbations
 in AdS$_4$. In order to use the AdS/CFT dictionary 
we need to write the AdS$_4$
metric in Fefferman-Graham coordinates 
\bea
\diff s^2 (\mathrm{AdS}_4)_{\mathrm{FG}} = \frac{1}{z^2}\left(\diff z^2   + \diff x_\mu \diff x^\mu\right)~,
\label{zads}
\eea
by changing  coordinates $\rho^2=1/z$. Here recall that $\rho$ is related asymptotically to $r$ via the 
change of variable (\ref{rchange}).
In particular, for scalar modes we then have
\bea
\varphi \sim \hat \varphi z^{-\Delta} + \varphi_0 z^{3-\Delta}~,
\label{vevis}
\eea
with $ \varphi_0$ corresponding to perturbing by an operator of dimension $\Delta$, and 
$\hat \varphi$ corresponding to the VEV of such an operator. 
Aided by the map between chiral multiplets in the SCFT and
 modes in the Kaluza-Klein spectrum on $V_{5,2}$, discussed earlier, 
we will see that the former possibility is realized. 

To see this, we examine the leading 
behaviour of the $G$-field at infinity, and the
corresponding pseudoscalar mode in AdS$_4$.  
We may discuss this in the context of general Sasaki-Einstein solutions and 
then specialize to the case of interest. Consider a self-dual harmonic
 $G$-flux in the Calabi-Yau cone background $\R^{1,2}\times C(Y)$, of the form
\bea
G = \alpha =  \diff ( \rho^{-\nu}\beta)~,
\label{homg}
\eea
where $\rho$ is the radial variable on the cone.
This implies 
$\Delta_Y \beta \,= \, \nu^2 \beta ~,$
where $\Delta_Y$ is the Laplace operator on $Y$ acting on three-forms. 
For the associated AdS$_4\times Y$ solution, we may then consider a fluctuation of the type $\delta C = \pi \cdot \beta$.
It was shown in \cite{Castellani:1984vv}  that  this leads to a pseudoscalar field $\pi$ in 
AdS$_4$ with mass\footnote{The reader should not confuse the mass $m^2$ here with the paramter $m$ in the deformed solution. }
\bea
m^2 \, =\,  \frac{\nu(\nu - 6 )}{4}~.
\eea
Substituting this into the formula for the dimension of the dual operator, $\Delta(\Delta - 3)= m^2$, 
 we obtain  $\Delta_\pm = \tfrac{1}{2}(3  \pm |3 -\nu  |)$. Which branch to pick depends {\it a priori}
 on the specific operator we consider.
 Going back to  our particular $G=\alpha$ given by (\ref{alpha}), 
we see  that 
\bea
\beta \, \propto \,  \left(3 \tilde\sigma_1 \wedge\tilde\sigma_ 2 \wedge\tilde \sigma_3 + \frac{1}{2}\epsilon_{ijk} \sigma_i \wedge \sigma_j \wedge \tilde \sigma_k \right)~, 
\eea
and $\nu = 4/3$. Then $\Delta_+ = 3 - \tfrac{\nu}{2} = \tfrac{7}{3}$, while $\Delta_- = \frac{\nu}{2} = \frac{2}{3}$.
Now, going through all the pseudoscalar modes undergoing shortening conditions 
 in the tables in \cite{Ceresole:1999zg}, we find a mode with $\Delta=\tfrac{7}{3}$ while the other possibility 
is not realized. In particular,  this mode arises as the
pseudoscalar component of the chiral operators with dimensions 
$\Delta = \tfrac{2}{3}m +1$, with $m=2$, that we discussed 
in section \ref{multipletsec}. 
From the asymptotic scaling $\alpha \sim z^{2/3}$, we conclude that this
operator is in fact added to the Lagrangian (see also \cite{Herzog:2000rz}).

Since this is the pseudoscalar component of a chiral superfield, we see that it is a Fermionic  mass term $\psi^\alpha \psi_\alpha$. This breaks parity invariance, which is reflected in the gravity solution in the presence of the internal flux, the latter being odd under 
parity.  In general, 
such mass terms may be added to the Lagrangian, in a supersymmetric way, by a quadratic 
\emph{superpotential} 
deformation\footnote{This deformation then introduces various additional terms in the Lagrangian. 
For example,  we have a quadratic term  $\mu^2 \mathrm{Tr}[\phi^\dagger\phi]$ 
in the bosonic F-term potential,  with dimension $\Delta=4/3$, as well as linear terms in $\mu$. 
Presumably these  operators may be detected by analysing appropriate
linearized perturbations of the background. However, their structure should be constrained by supersymmetry. See  \cite{Gomis:2008vc} for discussion of a related issue in the context of mass deformations of the ABJM theory.}
\bea
\delta W = \mu \mathrm{Tr}[\mathscr{\phi}^2]  ~~~~\Rightarrow ~~~~~ \delta \mathcal{L}=-\frac{1}{2}\frac{\de^2 \delta W}{\de {\phi}_i\de \phi_j}\psi_i^\alpha\psi_{j\, \alpha}+\ldots~.
\label{superpotdef}
\eea
{\it A priori}, we have three such possible mass terms, compatible with the $SU(2)_r$ global symmetry 
of the deformed background, namely
\bea
\delta  W = \frac{\mu_+}{2}\left(\mathrm{Tr}[\Phi_1^2]+\mathrm{Tr}[\Phi_2^2]\right)+
 \frac{\mu_-}{2} \left(\mathrm{Tr}[\Phi_1^2]-\mathrm{Tr}[\Phi_2^2]\right)+
\mu_3\mathrm{Tr}[A_1B_1+ A_2B_2]~.
\label{superpotdef2}
\eea
where in the above we mean superfields. 

We may deduce which terms are present by analysing more carefully the symmetries of the deformed solution. 
Recall from section \ref{csection} that in the undeformed field theory we have a $\zflip$
symmetry that exchanges $\Phi_1\leftrightarrow\Phi_2$, $A_i\leftrightarrow B_i$.  The generator acts 
on the $z_i$ coordinates, introduced just below equation (\ref{hypersurface}), as $(z_0,z_1,z_2,z_3,z_4 )\to (-z_0,z_1,-z_2,z_3,-z_4 )$. 
Hence $\zflip\subset  O(5)$ acts on the deformed quadric (\ref{deformation}).
The internal $G$-flux then \emph{breaks} this $\zflip$ symmetry. To see this, notice that for $k=1$ the 
zero-section of $\cX=T^* S^4$ is $S^4$, embedded in $\R^5$ by the real parts of the $z_i$ coordinates in 
(\ref{deformation}). The volume form on $S^4$ may be written
\bea
\vol(S^4)=\frac{1}{4!} \epsilon_{ijklm} z_i\diff z_j\wedge\diff z_k\wedge\diff z_l\wedge\diff z_m\mid_{\{\sum_{i=0}^4 z_i^2=\gamma^2~, \ \  z_i\in\R\}}~.
\eea
This hence changes sign under the generator of $\zflip$. Now since $\zflip$ is an isometry, it necessarily maps 
$L^2$ harmonic forms to $L^2$ harmonic forms, and as mentioned earlier the results of 
\cite{Hausel} imply that $G_{\mathrm{int}}\propto \alpha$ (\ref{solnn}), where $\alpha$ is given by (\ref{alpha}), is the only such form. Thus the generator 
of $\zflip$ maps $\alpha\mapsto\pm\alpha$. But since $\alpha$ restricts to the volume form on $S^4$ at $r=0$, 
we see that the generator of $\zflip$ maps $\alpha\mapsto -\alpha$, and thus $G_{\mathrm{int}}\mapsto -G_{\mathrm{int}}$. Hence the related superpotential deformation 
in (\ref{superpotdef2}) should also be odd. This requires that $\mu_+=\mu_3=0$, leaving precisely the 
following supersymmetric mass-term  
\bea\label{superdef}
W \to W +  \frac{\mu}{2} \left(\mathrm{Tr}[\Phi_1^2]-\mathrm{Tr}[\Phi_2^2]\right)~.
\eea
We may then regard the full superpotential as depending on the two parameters $s$ and $\mu$.
Notice that by setting $s=0$, the mass term $\mu$ is 
precisely that leading to the ABJM theory in the IR, 
after integrating out the adjoints. 

The deformed F-term equations following from the 
superpotential deformation (\ref{superdef}) read
\bea
\label{newFterms}
B_i\Phi_2 + \Phi_1 B_i  &=&0~,\\
\Phi_2 A_i + A_i\Phi_1  &=&0~,\\
3s\Phi_1^2 + (B_1A_1+B_2A_2) + \mu \Phi_1 &=&0~,\\
 3s\Phi_2^2 + (A_1B_1+A_2B_2) - \mu\Phi_2  &=&0~.
\eea
The simple linear change of variable
\bea
\Phi_1 = \Psi_1 - \frac{\mu}{6s}~,\qquad \Phi_2 = \Psi_2 + \frac{\mu}{6s}
\eea
then leads to 
\bea
\label{newerFterms}
B_i\Psi_2 + \Psi_1 B_i  &=&0~,\\
\Psi_2 A_i + A_i\Psi_1   &=&0~,\\
 3s\Psi_1^2 + (B_1A_1+B_2A_2)  &=& \frac{\mu^2}{12s}~,\\
 3s\Psi_2^2 + (A_1B_1+A_2B_2)   &=&\frac{\mu^2}{12s}~.
\eea
In particular, we see that the Abelian moduli space is exactly the deformed singularity 
(\ref{deformation}). The deformation parameter is proportional to the mass, $\gamma^2=\mu^2/12s$.

\subsection{Comments on the field theory in the IR}\label{sec:IR}

The supergravity solution implies that the ${\cal N}=2$ superconformal 
Chern-Simons-matter theory deformed by the mass term will flow in the IR to a 
\emph{confining} theory. 
We leave a field-theoretic understanding of this for future work, 
restricting ourselves here to making only some preliminary comments in this direction.  

Firstly, it is instructive to contrast the pattern of $U(1)_R$ symmetry breaking
of our solution with that of the Klebanov-Strassler theory. In the latter case 
the $U(1)_R$ symmetry is broken to $\Z_{2M}$ in the UV by the chiral anomaly, and this is then  
spontaneously broken to $\Z_2$, yielding  $M$ vacua.  On the gravity side, 
the breaking of  $U(1)_R$ to $\Z_{2M}$ is reflected by the non-invariance of the 
fluxes already in the UV \cite{Klebanov:2000nc,Klebanov:2002gr}. 
The $M$ vacua are then reflected by the presence of supersymmetric probe branes, representing
BPS domain walls interpolating between the vacua.
In three dimensions there is no chiral anomaly, and thus $U(1)_R$ cannot be broken in this way. 
Indeed, in the supergravity solution we discussed the parameter $M$ is \emph{not} a UV parameter 
that one can dial at infinity, and in fact the flux vanishes asymptotically. We also expect that 
no wrapped branes will give rise to BPS domain walls, although we have not checked this.

In analogy with the Klebanov-Strassler cascade, 
one possible way to interpret the RG flow described by the supergravity solution is to imagine that 
once the conformal theory is deformed by the mass term in the UV, it starts ``cascading'', going through a sequence of 
Seiberg-like dualities where the ranks of the gauge groups decrease, until in the deep IR perhaps one gauge group disappears, and the low energy-theory confines.
This idea has recently been suggested in \cite{Aharony:2009fc,Evslin:2009pk}
in the context of ABJM-like theories, although the models studied in these references
are different from our models. This interpretation is motivated by the brane creation mechanism that we discussed in section 
\ref{ftd}, and by the fact that in the solution there is a varying $B_2$-field (in the Type IIA reduction). More precisely, the $B_2$-field suggests that 
as we proceed to the IR, the NS5-branes rotate around the circle. 
Taking this point of view, and applying the duality rule of section \ref{ftd}, we end up in the IR
with a gauge group $U(-kM)_k\times U(kM)_{-k}$ after $M$ steps, which clearly doesn't make sense since one gauge group has negative rank. 
(We could of course stop applying the duality at the previous step.) Notice, however, that what is 
the precise gauge group in the IR depends on the starting point in the UV, which in turn depends on 
subleading corrections to $kM^2$.  In any case, it is not clear whether applying this rule is correct, 
once we turn on the mass deformation. In fact, more conservatively, given the mass term 
one should integrate out the heavy degrees of freedom, and obtain an effective 
low-energy theory in the IR. In principle this theory should then exhibit confinement (without supersymmetry breaking).
Integrating out the Fermions would {\it a priori} lead to a possible shift of the Chern-Simons levels. However, because the Fermions are in the adjoint representation in fact the levels are not shifted. Indeed, we have already noted that 
the mass term is exactly the same mass term which produces the ABJM theory at low energy, starting from the Chern-Simons
theory in Figure \ref{figv} with $s=0$. Integrating out the bosonic components of the chiral fields in the mass-deformed $n=2$ theory, the effective superpotential for the low-energy fields $A_i$, $B_i$ results in a non-local expression, involving square roots of polynomials in these fields. Hopefully, further work along these lines will lead to a precise identification of the IR field theory.

\section{Conclusions}
\label{concsec}

In this paper we have constructed a new example of AdS$_4/$CFT$_3$ duality by proposing a simple ${\cal N}=2$
Chern-Simons-matter quiver field theory as the holographic 
dual to the AdS$_4\times V_{5,2}/\Z_k$ Freund-Rubin background 
in M-theory. This duality presents several novel aspects. For example, the geometry, and hence the field theory, 
has an $SU(2)\times U(1)\times U(1)_R$ global symmetry (enhanced to $SO(5)\times U(1)_R$ for $k=1$), and hence these models 
are non-toric. Examples of AdS/CFT dual pairs of non-toric type, where both sides are known explicitly, are quite 
rare. This model may be thought of as describing the low-energy theory of multiple M2-branes at a quadric hypersurface
singularity. In fact, this is the $n=2$ member of a family of hypersurface singularities ($\mathcal{A}_{n-1}$ four-fold singularities), 
labelled by a positive integer $n$, for which we have also presented the corresponding field theories. However, we have explained that only for $n=2$ and $n=1$ do these singularities give rise to Freund-Rubin AdS$_4$ duals, the $n=1$ model being the ABJM theory. We note that 
\cite{Gauntlett:2006vf} discussed the larger class of ADE four-fold singularities, and it was shown in this reference that 
in this class the only cases that can admit Ricci-flat K\"ahler cone metrics are $\mathcal{A}_0=\C^4$, $\mathcal{A}_1$ and $\mathcal{D}_4$. It would be interesting 
to construct Chern-Simons-matter theories dual to other hypersurface singularities, 
and to see whether the $\mathcal{D}_4$ theory admits 
a Freund-Rubin holographic dual, analogous to that discussed in this paper. 

In this paper we have considered the case where the Chern-Simons levels are equal $k_1=-k_2=k$. 
Relaxing this condition, thus allowing for arbitrary levels, corresponds to deforming 
the Type IIA solutions that we discussed in  section \ref{redux}
by turning on a Romans mass \cite{alessandro}. Such solutions will be similar to 
those discussed in \cite{Petrini:2009ur,Lust:2009mb} and it would 
be interesting to find these solutions explicitly.

Another interesting aspect of the model we discussed is that there exists a deformed supergravity solution, 
that we have argued corresponds to 
a particular supersymmetric mass deformation of the conformal theory. This deformation is similar to those studied in \cite{Hosomichi:2008qk,Hosomichi:2008jd,Gomis:2008vc} and other 
references. 
We have seen that this mass term is dual to a harmonic $(2,2)$, primitive (hence self-dual) 
$G$-flux on the Calabi-Yau geometry. Quite recently the authors of reference \cite{Lambert:2009qw} have shown how self-dual background fluxes 
induce mass terms in the M2-brane worldvolume action, and it would be interesting to see whether this construction generalizes to
 ${\cal N}=2$ backgrounds of the type we have studied.  
In the present context the effect of this mass term 
is rather different from that in the ABJM model studied in \cite{Hosomichi:2008qk,Hosomichi:2008jd,Gomis:2008vc}:
it deforms the classical moduli space in a way that precisely 
matches the geometry in the supergravity dual. 
In particular, the solution develops a finite-sized $S^4$ in the IR, implying that 
the theory becomes confining.  Motivated by brane constructions, 
we have briefly discussed how this deformation might be interpreted as a ``cascade'', 
analogous to the Klebanov-Strassler cascade. 
However, further work is needed in order to obtain a more conclusive interpretation of the RG flow, 
and in particular a clearer understanding of the field theory in the deep IR. 
We expect a similar story to repeat for other deformed 
solutions with self-dual $G$-flux, based on different special holonomy 
manifolds \cite{Cvetic:2001zb,Cvetic:2000db}. 

Finally, in appendix \ref{mirror} we describe a 
Type IIA reduction of the supergravity solutions that is \emph{different} 
to that considered in the main text, {\it i.e.} we reduce on a different choice of M-theory circle. 
On general grounds, one expects this to 
lead to a field theory that is mirror to that considered in section 
\ref{sec2} (see, for example, \cite{Gukov:2002es}). It would be interesting to study this reduction 
further.

\subsection*{Acknowledgments}

We are very grateful to A.~Armoni, S.~Gukov, P.~Kumar, J.~Maldacena, C.~Nu\~nez, 
M.~Piai, Y.~Tachikawa and D.~Tong for discussions. J.~F.~S. is supported by a Royal 
Society University Research Fellowship.

\appendix 

\section{Some cohomology computations}
\label{apptop}

In the main text we have made use of a number of different cohomology groups of the various manifolds we have defined, 
and also the relations between the groups. In this appendix we present the relevant computations.

\

We begin by defining a manifold that does \emph{not} appear in the main text: we define 
$\cX_n$ by
\bea\label{deformXn}
\cX_n = \left\{\prod_{\gamma=1}^n (z_0-a_\gamma) + \sum_{i=1}^4 z_i^2=0\right\}\subset\C^5~.
\eea
Here the $a_\gamma$, $\gamma=1,\ldots,n$, are real, pairwise non-equal constants, which we order as 
$a_1<a_2<\cdots<a_n$. The manifold $\cX_2=\cX$ in the main text, which is the deformation of the quadric 
singularity. The $\cX_n$ are smooth non-compact complex manifolds with boundaries 
$\partial \cX_n=Y_n$, where $Y_n$ is defined by (\ref{nfourfolds}), (\ref{base}). 
Indeed, the $\cX_n$ are \emph{deformations} of the $X_n$ singularities (\ref{nfourfolds}). 

The cohomology of $\cX_n$ was discussed in \cite{Gukov:1999ya}, and we briefly review their analysis. 
For $\gamma=1,\ldots,n-1$ we may define a four-sphere $S^4_\gamma$ by requiring
that $z_0$ is real with $a_\gamma<z_0<a_{\gamma+1}$, and that the $z_i$, for $i=1,\ldots,4$, are all real or all imaginary,
depending on the value of $\gamma$ mod 2. These $n-1$ four-spheres then generate 
$H_4(\cX_n,\Z)\cong \Z^{n-1}\cong H^4(\cX_n,Y_n,\Z)$, where the last 
step is Poincar\'e-Lefschetz duality. This is the only non-trivial homology group 
of $\cX_n$ (of course $H_0(\cX_n,\Z)\cong\Z$). Each four-sphere has self-intersection 
number 2, since its normal bundle may easily be seen to be $T^*S^4$ which has Euler number 
2, and by construction the intersection number of $S^4_\gamma$ with $S^4_{\gamma+1}$ is 
1, with all other intersection numbers vanishing. Poincar\'e-Lefschetz duality 
implies that $H^4(\cX_n,Y_n,\Z)$ and $H^4(\cX_n,\Z)$ are dual lattices, where 
recall that $f:H^4(\cX_n,Y_n,\Z)\rightarrow H^4(\cX_n,\Z)$ forgets that a class is 
relative (has compact support). Thus the above discussion shows that 
$H^4(\cX_n,Y_n,\Z)\cong H_4(\cX_n,\Z)$, equipped with the intersection form, is the 
root lattice of $\mathcal{A}_{n-1}$, while $H^4(\cX_n,\Z)$ is the dual weight lattice. 

Notice that in the simple case with $n=2$, where $\cX_2=\cX\cong T^*S^4$, 
the generator of $H^4(\cX_2,Y_2,\Z)\cong\Z$ may be taken to be a compactly supported 
four-form that has integral one over the fibre (the Thom class of the bundle $T^*S^4$).

\

We may now compute the cohomology of $Y_n=\partial \cX_n$ using the long exact sequence 
for the pair $(\cX_n,Y_n)$. Since the cohomology groups of both $\cX_n$ and $(\cX_n,Y_n)$ 
vanish in all degrees other than the top, middle and bottom, it follows that 
most of the cohomology of $Y_n$ is also trivial. In fact the only non-trivial 
cohomology group is $H^4(Y_n,\Z)$, which arises from the sequence
\bea
\cdots \longrightarrow H^4(\cX_n,Y_n,\Z) \stackrel{f}{\longrightarrow} H^4(\cX_n,\Z)\longrightarrow H^4(Y_n,\Z)\longrightarrow 
H^5(\cX_n,Y_n,\Z)\cong 0~.
\eea
This implies that $H^4(Y_n,\Z)\cong H^4(\cX_n,\Z)/f(H^4(\cX_n,Y_n,\Z))\cong \Z_{n}$, 
where the last isomorphism follows from the above description of the cohomology groups in terms of the 
root and weight lattices of $\mathcal{A}_{n-1}$. Of course, by Poincar\'e duality we also have 
$H_3(Y_n,\Z)\cong\Z_n$.

In the special case that $n=2$, of main interest in the text, the long exact homology sequence 
implies that we may take the boundary $S^3$ of any fibre $S^3=\partial \R^4$ of $T^*S^4$ as 
the generator of $H_3(Y_2,\Z)\cong\Z_2$. Equivalently, viewing $Y_2$ as an $S^3$ bundle over 
$S^4$, a copy of the fibre at any point on the base generates this third homology group.

\

Next we introduce the free circle action on $Y_n$ by $U(1)_b\cong SO(2)_{\mathrm{diag}}\subset SO(4)$, 
where $SO(4)$ acts on the coordinates $z_i$, $i=1,\ldots,4,$ in the vector representation. 
The quotient $M_n=Y_n/U(1)_b$ is then a smooth compact six-manifold. The cohomology of this 
space may be deduced from the Gysin sequence for the circle fibration of $Y_n$ over $M_n$:
\bea\label{gysin}
\cdots\longrightarrow && H^{i-2}(M_n,\Z)\stackrel{\cup c_1}{\longrightarrow} H^i(M_n,\Z)
\longrightarrow H^i(Y_n,\Z)\longrightarrow \nonumber\\
&& H^{i-1}(M_n,\Z)\longrightarrow\cdots~.
\eea
It is straightforward to derive this sequence from the long exact sequence 
for the total space $\mathcal{L}$ of the complex line bundle over $M_n$ associated 
to the $U(1)_b$ circle bundle: note that $\mathcal{L}$ has boundary $Y_n$, 
and base $M_n$. One needs to combine this sequence with the Thom 
isomorphism -- this is precisely where the cup product with $c_1=c_1(\mathcal{L})$ 
comes from above, since for a complex line bundle $c_1$ is equal to the Euler class 
of the underlying rank 2 real vector bundle. The last map in the Gysin sequence 
(\ref{gysin}) is just pull-back from $M_n$ to $Y_n$.

Using the sequence (\ref{gysin}), together with the known cohomology of $Y_n$ 
computed above, we may compute the cohomology (and properties 
of the cohomology ring) of $M_n$. Since $H^1(Y_n,\Z)\cong H^2(Y_n,\Z)\cong 0$, it follows immediately 
from $i=2$ in (\ref{gysin}) that $c_1\equiv\Omega_2$ is the generator of $H^2(M_n,\Z)\cong\Z$. 
Here the notation $\Omega_2$ was introduced in the main text just before equation (\ref{thebfield}).
Similarly, $H^3(Y_n,\Z)\cong 0$ implies that $H^3(M_n,\Z)\cong 0$. 
Then $i=4$ above implies $\Z_n\cong H^4(Y_n,\Z)\cong H^4(M_n,\Z)/[H^2(M_n,\Z)\cup c_1]$. 
Now, $H^4(M_n,\Z)\cong H_2(M_n,\Z)$, so the free part of $H^4(M_n,\Z)$ is 
$\Z\cong H^2(M_n,\Z)$ by the Universal Coefficient Theorem. 
Moreover, the torsion in $H^4(M_n,\Z)$ is the torsion in $H_3(M_n,\Z)$, 
but this is Poincar\'e dual to $H^3(M_n,\Z)\cong0$. Thus 
$H^4(M_n,\Z)\cong \Z$, and the Gysin sequence thus 
tells us that the square of the generator of 
$H^2(M_n,\Z)$ is \emph{$n$ times} the generator of $H^4(M_n,\Z)$. 
We may equivalently state this as
\bea\label{square}
\int_{\Sigma^4} \Omega_2\cup \Omega_2= n~,
\eea
where $\Sigma^4$ denotes the generator of $H_4(M_n,\Z)$, again as in the main text. 
The result (\ref{square}) follows from Poincar\'e duality, and 
the last map in the Gysin sequence that says 
cupping $H^4(M_n,\Z)$ with $c_1=\Omega_2$ (which is Poincar\'e dual to $\Sigma^4$) maps 
the generator of $H^4(M_n,\Z)$ to the generator of $H^6(M_n,\Z)\cong\Z$. Notice 
that $M_n$ then has the same cohomology groups as $\mathbb{CP}^3$ (where $M_1\cong \mathbb{CP}^3$), 
but that the cohomology \emph{ring} depends on $n$ via the above calculation.

\

We may now compute the cohomology of the quotient $\ymodk$. This is 
also a smooth seven-manifold, where we take $\Z_k\subset U(1)_b$. 
This immediately gives $\pi_1(\ymodk)\cong H_1(\ymodk,\Z)\cong \Z_k$. The 
Gysin sequence (\ref{gysin}), with $\ymodk$ in place of $Y_n$, now 
has $c_1=k\Omega_2$. Precisely as we argued above, this implies the important result that
 $H^4(\ymodk,\Z)\cong H^4(M_n,\Z)/[H^2(M_n,\Z)\cup k\Omega_2]\cong \Z_{nk}$. 
Of course, by Poincar\'e duality also $H_3(\ymodk,\Z)\cong\Z_{nk}$. 
Indeed, the Poincar\'e dual sequence implies that the generator 
$\Sigma^2$ of $H_2(M_n,\Z)\cong\Z$ lifts to the generator 
$\Sigma^3$ of $H_3(\ymodk,\Z)\cong\Z_{nk}$, where $\Sigma^3$ 
is the total space of the circle bundle over a representative 
of $\Sigma^2$. This was used at the end of section \ref{IIAsec}.

Finally, recall that in the special case of $n=2$ the generator 
of $H_3(Y_2,\Z)\cong\Z_2$ can be taken to be a copy of 
the fibre $S^3$ in the fibration $S^3\hookrightarrow Y_2\rightarrow S^4$. 
The fibres over the poles $p_N$, $p_S$ of the $S^4$ are mapped into themselves 
under $\Z_k$, with the Hopf action of $\Z_k$ on $S^3$ giving the quotient 
$S^3/\Z_k$. It then follows from the last paragraph that this Lens 
space $S^3/\Z_k\cong \Sigma^3$ generates $H_3(Y_2/\Z_k,\Z)\cong\Z_{2k}$.

\section{The Stenzel metric}
\label{appdetails}

In this appendix we review the construction of the Stenzel metric on $\cX\cong T^* S^4$.
The deformed quadric $\cX$ is defined as
\bea
\sum_{i=0}^4 z_i^2 = \gamma^2~,
\label{singsing}
\eea
and the Stenzel metric on this may be written by introducing left-invariant one-forms $L_{AB}$ on $SO(5)$, 
$A,B=1,\ldots,5$, satisfying $\diff L_{AB}=L_{AC}\wedge L_{CB}$.
We split $A=(1,2,i)$, with $i=1,2,3$, where the $L_{ij}$ 
are left-invariant one-forms for $SO(3)$, and define
\bea
\sigma_i=L_{1i}, \qquad \tilde{\sigma}_i=L_{2i}, \qquad 
\nu = L_{12}~.
\eea
These are one-forms on the coset space $V_{5,2}=SO(5)/SO(3)$. 
The metric on (\ref{singsing}) is then \cite{Cvetic:2000db}
\bea
\diff s^2 = c^2 \diff r^2 + c^2 \nu^2 + a^2 \sigma_i^2 + b^2 \tilde{\sigma}_i^2~.
\eea
It is useful to introduce the orthonormal frame
\bea
e^0 = c\diff r~,\qquad e^{\tilde{0}}=c\nu~,\qquad e^i = a\sigma_i~,\qquad e^{\tilde{i}}=b\tilde{\sigma}_i~.
\label{frame}
\eea
A holomorphic frame is provided by 
\bea
\epsilon^0 = -e^0+\mathrm{i}e^{\tilde{0}}~,\qquad \epsilon^i = e^i+\mathrm{i}e^{\tilde{i}}~.
\eea
In this frame, we take the K\"ahler form $J$ and holomorphic $(4,0)$-form $\Omega$ 
to be the standard forms
\bea
J=\frac{\mathrm{i}}{2}\epsilon^\alpha\wedge \bar{\epsilon}^{\bar{\alpha}}, \qquad \Omega=\epsilon^0\wedge\epsilon^1\wedge 
\epsilon^2\wedge\epsilon^3~.
\eea
Thus these automatically satisfy the $SU(4)$-structure algebraic relations $J\wedge\Omega=0$, $\frac{1}{4!}J^4 = \frac{1}{16}\Omega\wedge
\bar{\Omega} = -e^{0\tilde{0}1\tilde{1}2\tilde{2}3\tilde{3}}$. A Ricci-flat K\"ahler metric requires 
$\diff J=0=\diff\Omega$. It is straightforward to check that 
$\diff J=0$ is equivalent to the ordinary differential equation (ODE)
\bea
(ab)'=c^2~,
\eea
where a prime denotes differentiation with respect to $r$, while imposing $\diff\Omega=0$ is equivalent to the four ODEs
\bea
3\frac{a'}{a}+\frac{c'}{c}-3\frac{b}{a}&=&0~,\nn\\
3\frac{b'}{b}+\frac{c'}{c}-3\frac{a}{b}&=&0~,\nn\\
2\frac{a'}{a}+\frac{b'}{b}+\frac{c'}{c}-2\frac{b}{a}-\frac{a}{b}&=&0~,\nn\\
2\frac{b'}{b}+\frac{a'}{a}+\frac{c'}{c}-2\frac{a}{b}-\frac{b}{a}&=&0~.
\eea
Although this naively looks overdetermined, it is simple to check by taking linear combinations that these five ODEs are equivalent to the three ODEs
\bea\label{firstorderODEs}
\frac{a'}{a}&=&\frac{b^2+c^2-a^2}{2ab}~,\nn\\
\frac{b'}{b}&=&\frac{a^2+c^2-b^2}{2ab}~,\nn\\
\frac{c'}{c}&=&\frac{3(a^2+b^2-c^2)}{2ab}~.
\eea
This is the same system of equations that were presented in \cite{Cvetic:2000db}, although in the latter reference 
they were derived by first finding the second order Einstein equations, and then constructing a superpotential. 
Here we have derived them directly from the Ricci-flat K\"ahler conditions.
A solution to these equations, which is a smooth complete metric on $\cX=T^*S^4$, was found by Stenzel \cite{stenzel}.
This is the solution written in (\ref{stenzsol}).

\section{A different reduction to Type IIA}\label{mirror}

In sections \ref{redux} and \ref{IIAsec} we considered reducing M-theory on $\R^{1,2}\times X_n/\Z_k$ with $N$ spacefilling M2-branes, or 
its near-horizon limit AdS$_4\times \ymodk$, along $U(1)_b$ to Type IIA string theory. Recall here that $X_n$ admits a Ricci-flat K\"ahler cone 
metric only for $n=1$ and $n=2$. In the case $n=2$, one problem with this Type IIA reduction is 
that as soon as one deforms the AdS$_4\times Y_2/\Z_k$ 
solution to the $\mathbb{R}^{1,2}\times \cX_2/\Z_k$ solution, the reduction along $U(1)_b$ is no longer well-behaved. 
Specifically, the $U(1)_b$ action fixes the north and south poles of the $S^4$ zero-section of $\cX\equiv\cX_2\cong T^*S^4$; since these 
are codimension eight, there is no simple interpretation of the resulting singularity in the dilaton in Type IIA string theory. 
Thus the Type IIA supergravity solution cannot be trusted in the IR region near to the $S^4$ at $r=0$.
However, there is a \emph{different} reduction to Type IIA that \emph{is} well-behaved. We briefly describe this here,
leaving a more thorough investigation for future work. 

Recall that in section \ref{secRRflux} we introduced a different $U(1)\equiv U(1)_6$ action on $X_n$. If we regard $X_n$ as being defined by the 
hypersurface equation (\ref{hypersurface}), the coordinates $(A_1,A_2,B_1,B_2,z_0=[s(n+1)]^{1/n} \Phi_2)$ have charges $(1,0,-1,0,0)$ under $U(1)_6$. 
In fact, we may \emph{deform} $X_n$ to $\cX_n$ given by (\ref{deformXn}), so that $U(1)_6$ also acts on the smooth 
manifold $\cX_n$. Of course, to obtain a solution to eleven-dimensional supergravity, we should equip $\cX_n$ with a 
Calabi-Yau metric. For $n=1$, $n=2$, we may use complete asymptotically conical Calabi-Yau metrics (the flat metric on $\cX_1\cong\C^4$; the 
Stenzel metric on $\cX_2\cong T^*S^4$). These are the metrics relevant for application to the AdS/CFT correspondence. Such metrics do not exist for $n>2$, 
in which case the reader can imagine that (\ref{deformXn}) is a local model in a \emph{compact} Calabi-Yau manifold. 
Yau's theorem will then give a Ricci-flat K\"ahler metric on this space which is incomplete at the boundary. In any case, the 
precise details of the metric will not be important in what follows.

Consider reduction of M-theory on $\R^{1,2}\times \cX_n$, with $N$ spacefilling M2-branes, along $U(1)_6$. 
The fixed point set is codimension four, namely $\{A_1=B_1=0\}$, which cuts out the locus
\bea\label{An}
\prod_{\gamma=1}^n (z_0-a_\gamma) + A_2B_2=0~.
\eea
This is the deformation of the $\mathcal{A}_{n-1}$ singularity: it has $n-1$ two-spheres $S^2_\gamma$, defined 
similarly to the four-spheres $S^4_{\gamma}$ in appendix \ref{apptop},
that intersect according to the root lattice of $\mathcal{A}_{n-1}=SU(n)$. 
This becomes a D6-brane locus when we reduce to Type IIA. Indeed, 
the Type IIA spacetime is flat, since $\cX_n/U(1)_6\cong \R^7$. 
To see this, note that $\cX_n/\C^*_6$ is described by 
\bea
z + \prod_{\gamma=1}^n (z_0-a_\gamma) + A_2B_2=0~.
\eea
where $z=A_1B_1$. This is simply $\C^3$. 
The quotient space is thus diffeomorphic to $\R^7\cong\R_7\times \C^3$, 
where $\R_7$ is spanned by $|A_1|^2-|B_1|^2$, which one can think of as the moment map 
for $U(1)_6$, and $\C^3$ is spanned by $(A_2,B_2,z_0)$. The fixed point locus 
is thus at the origin of $\R_7$, and cuts out the hypersurface 
(\ref{An}) in the $\C^3$ part. 

The reduction of $\R^{1,2}\times \cX_n$ along $U(1)_6$ is thus 
the flat spacetime $\R^{1,9}=\R^{1,2}\times \R_7\times\C^3$, with 
$N$ spacefilling D2-branes and a single spacefilling D6-brane 
sitting at the origin of $\R_7$ and wrapping the divisor
(\ref{An}) in $\C^3$. Notice that this 
description gives the correct amount of supersymmetry, 
since a D-brane wrapped on a divisor in a three-fold 
preserves four supercharges, or $\mathcal{N}=2$ supersymmetry
in $d=3$.  

There are $n-1$ four-cycles in $\cX_n$, and the quantized $G$-flux through the 
generators $S^4_\gamma$ defined in appendix \ref{apptop} gives 
\bea
\frac{1}{(2\pi l_p)^3} \int_{S^4_\gamma} G = {M}_\gamma\in\mathbb{Z}~.
\eea
In the Type IIA reduction considered in this section, this is dual 
to adding ${M}_\gamma$ units of \emph{worldvolume} gauge field 
flux on the D6-brane through the two-sphere $S^2_\gamma$ in the deformed 
$\mathcal{A}_{n-1}$ singularity (\ref{An}). A general discussion of this may be found in \cite{Sparks:2003ck}. Thus
\bea
\frac{1}{2\pi l_sg_s}\int_{S^2_\gamma} F = {M}_\gamma~,
\eea
where $F$ is the $U(1)$ gauge field on the D6-brane.

In the limit where $a_\gamma\rightarrow 0$, which is the hypersurface singularity $X_n$, the D6-brane is wrapped on 
$\R^{1,2}\times \mathcal{A}_{n-1}$ (we emphasize that the \emph{spacetime} is flat Minkowski spacetime). In particular, for $n=2$ we have an $\mathcal{A}_1$ singularity, although 
for $n>2$ the above analysis shows that the $\mathcal{A}_1$ \emph{quiver} in section \ref{sec2} 
is \emph{not} related to this $\mathcal{A}_1$ singularity in the Type IIA reduction on $U(1)_6$. 
Indeed, since we are reducing on a different circle, one expects the effective gauge theory 
derived from the brane configuration described above to be \emph{mirror} to the 
gauge theory in section \ref{sec2}, which we derived from the Type IIA reduction on 
$U(1)_b$ in section \ref{IIAsec}. 

We may also consider taking the $\Z_k$ quotient along $U(1)_b$. The charges of the coordinates $(A_1,A_2,B_1,B_2,z_0)$ 
under $U(1)_b$ are $(1,1,-1,-1,0)$, and thus in the Type IIA internal space 
$\R_7\times\C^3$, spanned by the moment map $|A_1|^2-|B_1|^2$ and $(A_2,B_2,z_0)$, respectively, 
$U(1)_b$ acts with charges $(1,-1,0)$ on $\C^3$. Thus the $\Z_k$ quotient along $U(1)_b$ 
leads to a \emph{$\Z_k$ singularity in spacetime}, or more precisely 
an $\mathcal{A}_{k-1}$ singularity. This would usually lead to an $SU(k)$ gauge symmetry 
in the transverse six-dimensional space. Contrast this with the $\mathcal{A}_{n-1}$ singularity 
on which the D6-brane is wrapped.

Finally, notice that we may perform a T-duality along the $U(1)$ which acts with 
charges $(1,-1)$ on the coordinates $(A_2,B_2)$.  This gives a Type IIB brane set-up 
where the spacetime is $\R^{1,2}\times \R_7\times S^1\times \R^5$, with $N$ 
spacefilling D3-branes wrapping the $S^1$ circle (that arises from the T-duality). 
Here $\R^5$ arises as $\R^5=\R\times\C^2$, where $\R$ is spanned by the moment map 
$|A_2|^2 - |B_2|^2$, and $\C^2$ is spanned by $(z_0,A_2B_2)$. 
Since the fixed point locus is $\{A_2=B_2=0\}$, which is a copy of $\R^{1,2}\times\R_7\times 
\C$ in the IIA spacetime (with $\C$ spanned by the coordinate $z_0$), on T-dualizing this becomes a 
linearly embedded spacefilling NS5-brane. More precisely, the NS5-brane 
wraps the $\R_7$ direction, sits at a point in $S^1$, and wraps the copy 
of $\C\subset\R^5$ spanned by the coordinate $z_0$. When we divide by $\Z_k\subset U(1)_b$, 
the fixed locus is precisely the $\mathcal{A}_{k-1}$ singularity, and we thus obtain 
$k$ linearly embedded spacefilling NS5-branes in the Type IIB dual.
The spacefilling D6-brane wrapped on the deformation 
of the $\mathcal{A}_{n-1}$ singularity becomes a spacefilling D5-brane wrapped on a 
non-linearly embedded copy of $\R^3$ in $\R^5$. This is because the four-manifold (\ref{An}) 
fibres over $\R^3$ with $n$ fixed points. 
The two copies of $\R^3$ wrapped by the D5-brane and the $k$ NS5-branes thus 
intersect at $n$ points in $\R^6=\R_7\times \R\times\C^2$.

\end{document}